\documentclass[twocolumn, showpacs, preprintnumbers,superscriptaddress, amsmath, amssymb, pre]{revtex4-2}
\usepackage{graphicx}
\usepackage{dcolumn}
\usepackage{bm}
\usepackage{epstopdf}

\usepackage{color}

\begin{document}

\title{
Self-Consistent Random Phase Approximation from Projective Truncation Approximation Formalism}

\author{Yue-Hong Wu}
\affiliation{School of Physics, Renmin University of China, 100872 Beijing, China}
\affiliation{Key Laboratory of Quantum State Construction and Manipulation (Ministry of Education), Renmin University of China, Beijing 100872, China}

\author{Xinguo Ren}
\email{renxg@iphy.ac.cn}
\affiliation{Institute of Physics, Chinese Academy of Sciences, Beijing 100190, China}

\author{Ning-Hua Tong}
\email{nhtong@ruc.edu.cn}
\affiliation{School of Physics, Renmin University of China, 100872 Beijing, China}
\affiliation{Key Laboratory of Quantum State Construction and Manipulation (Ministry of Education), Renmin University of China, Beijing 100872, China}

\date{\today}

\begin{abstract}

We derive the self-consistent random phase approximations (sc-RPA) from the projective truncation approximation (PTA) for the equation of motion of two-time Green's function. The obtained sc-RPA applies to arbitrary temperature and recovers the Rowe's formalism at zero temperature. The PTA formalism not only rationalize Rowe's formula, but also provides a general framework to extend sc-RPA. We implement the sc-RPA calculation for the one-dimensional spinless fermion model in the parameter regime of disordered ground state, with the N-representability constraints enforced. The obtained ground state energy, correlation function, and density spectral function agree well with existing results. The features of the Luttinger liquid ground state and the continuum/bound state in the spectral function are well captured. We discuss several issues concerning the approximations made in RPAs, difficulties of RPA for symmetric state, and the static component problem of PTA.

\end{abstract}


\maketitle

\begin{section}{Introduction}

Random phase approximation (RPA), as a theoretical method for studying properties of excited as well as ground state of correlated fermion systems, is widely used in solid state physics, nuclear physics, and quantum chemistry. Since its first appearance \cite{Pines1,Pines2,Pines3,Pines4}, 
RPA (in a variety of forms) has been derived in different ways, such as the the equation of motion (EOM) formalism of Rowe \cite{Rowe1}, the summation of bubble diagrams in the diagrammatic perturbation theory \cite{Hubbard1,Li4}, time-dependent Hartree approximation \cite{Ehrenreich1}, time-dependent density matrix approach \cite{Schuck4}, and the functional derivation method \cite{Hedin1,Giuliani1}. These approaches differ not only in the form of the resulting equations, but also in the aspects such as applicability in zero temperature/finite temperature, canonical ensemble/grand canonical ensemble, and the extendibility to higher order correlations. Among them, Rowe's EOM formalism of RPA \cite{Rowe1} has the advantage that excitation energies and the ground state properties can be obtained directly by solving a generalized eigenvalue problem. Higher order correlations are also easy to include. It is therefore widely used in nuclear study and quantum chemistry calculation. 
For comparable studies of RPA and its variants, see Refs.\cite{Co1,Li3}.

It is worthwhile to mention that, in the field of computational chemistry and materials science, RPA has been developed into a powerful first-principles method for real molecules and materials \cite{Hesselmann1,Eshuis1,Ren1,Chen2017,Zhang/Ren:2025}. In this context, RPA was formulated as a fifth-rung exchange-correlation functional \cite{Perdew/Schmidt:2001,Zhang/Ren:2025} of density functional theory \cite{Kohn1} via the framework of adiabatic connection fluctuation dissipation theorem \cite{Langreth1,Gunnarsson1}. Compared to the local and semilocal approximations of the XC functional, RPA can capture non-local 
electron correlations, allowing for the description of, e.g., van der Waals interactions \cite{Dobson1,Ren3}. In addition, it has been shown \cite{Scuseria1} that RPA is intimately connected to coupled cluster theory \cite{Bartlett1}, a hierarchical  series of methods that reaches the gold standard of main-group chemistry. While originating from the same basic concept,
the actual formulation and practical aspects of RPA for first-principles calculations and for solving correlated lattice models are rather different. Exploring the links and 
differences of RPA used in these two context, especially regarding self-consistent RPA formalism, is a highly interesting subject. For this purpose, a general and extendable formalism of RPA is always desirable.

In this work, we derive a self-consistent version of RPA from the projective truncation approximation (PTA) \cite{Fan1}. PTA is similar in spirit to Rowe's EOM method but differs in some important aspects. In Rowe's original idea, for systems whose low lying states form a harmonic space, by analogy to the non-interacting boson systems, the approximate excitation operator $O^{\dagger}$ is considered to satisfy $[H, O^{\dagger}] = \omega O^{\dagger} +P$, where $P$ applies outside the harmonic space. The low energy excitation energy $\omega$ is obtained from the equation
\begin{equation}    \label{revise_eq1}
   \langle \phi | [R, [H, O^{\dagger}]]| \phi \rangle = \omega \langle \phi| [R, O^{\dagger}]| \phi \rangle,
\end{equation}
where $R$ is an arbitrary operator and $| \phi \rangle$ lies within the harmonic space. RPA is obtained by expressing both $O^{\dagger}$ and $R$ in linear combinations of particle-hole excitation operators and carrying out linear variations to $R$ \cite{Rowe1}. An equivalent formulation is to minimize the normalized energy-weighted sum rule \cite{Schuck1}
\begin{equation}
   \omega = \frac{1}{2} \frac{\langle 0 | [Q, [H, Q^{\dagger}]] | 0 \rangle }{\langle 0 | [Q,  Q^{\dagger}] | 0 \rangle }
\end{equation}
with respect to the operator $Q$, which is the linear combination of particle-hole excitation operators.
In both formalisms, the states $|\phi \rangle$ and $|0\rangle$ take the approximate ground state.

In PTA, the chains of EOM for the two-time Green's functions (GFs) are truncated by the operator projection method developed by Zwanzig \cite{Zwanzig1}, Mori \cite{Mori1,Mori2}, Tserkovnikov \cite{Tserkovnikov1,Tserkovnikov2}, {\it etc.} in 1960s. In the form developed by Fan et al. \cite{Fan1}, PTA has been shown to be a systematic, well-controlled, and practical method for studying quantum \cite{Fan2,Fan3} as well as classical \cite{Ma1,Jia1} many-body systems. Our previous work on PTA focused on the single-particle GFs \cite{Fan1,Ma2}. We now apply PTA to two-particle GFs and this naturally leads to the self-consistent RPA (sc-RPA). 

This new formalism differs from that of Rowe's EOM approach \cite{Rowe1} in the following key aspects. In contrast to Rowe's idea where the operator projection is implicitly used, in PTA, the concept of operator projection is explicitly employed. By introducing the inner product $(X|Y)$ between two arbitrary operators $X$ and $Y$, we can assign a length $|X| \equiv \sqrt{|(X|X)|}$ to $X$. Rowe's idea can now be regarded as minimizing the difference between $[H, O^{\dagger}]$ and $\omega O^{\dagger}$, i.e., minimizing $| [H, O^{\dagger}] - \omega O^{\dagger} |$, with respect to the operator $O^{\dagger}$ under the commutator inner product $(X|Y) = \langle 0 | [X^{\dagger}, Y] |0 \rangle$. This is equivalent to the problem of projecting $[H, O^{\dagger}]$ to $O^{\dagger}$ in the operator space under a special inner product. By formally introducing the operator projection and allowing for more general inner product, the PTA-based formalism not only rationalizes Rowe's approach, but also provides a flexible framework to systematically extend RPA to higher order correlations. While Rowe's EOM formalism can only produce ground as well as excited state energies and the corresponding transition matrix elements, the PTA-based RPA can produce both the static averages and the dynamical spectral functions at arbitrary temperatures. This formalism also allows us to develop possible variants of RPA by considering different inner product definitions.

In this work, we also make a thorough comparison of the present formalism with previous EOM-based RPA. For the number operator method for calculating density matrix, we refer mainly to the work of Rowe \cite{Rowe2}. For the sc-RPA scheme, we refer to the work of Schuck's group \cite{Schuck1}. For the more detailed formula for higher order correlation at $T=0$, we refer to the work of Catara \cite{Catara1}. For demonstration purposes, we apply our sc-RPA to the one-dimensional interacting spinless fermion models. The results for ground state energy, occupation of fermions in momentum space, correlation function, and the spectral function are compared with the results from exact daigonalization, Bethe Ansatz and bosonization. Good agreement is achieved in the weak to intermediate interaction regime. Some key features of the Luttinger liquid ground state and excitations are captured in the large size limit, showing the importance of the self-consistent calculation in describing the correlation effect.

The plan of this paper is as follows. We follow a general-to-specific way so that it is easier to clarify what approximations are introduced at each stage. In Sec.II, we first present the general PTA for the EOM of bosonic GF. In Sec.III, we choose a special set of operators as our basis, namely the set of one-body density operators of the form $a_{\alpha}^{\dagger} a_{\beta}$. Here, $a_{\alpha}^{\dagger}$ and $a_{\beta}$ are creation and annihilation operators of fermions on the single-particle orbitals $\alpha$ and $\beta$, respectively ($1 \leq \alpha \neq \beta \leq L$, $L$ being the number of single-particle orbitals). This will lead to the formula of the general sc-RPA. We discuss the self-consistent calculation of one- and two-body density matrices at arbitrary temperature in this framework. Comparison with Rowe's formalism is made. In Sec.IV, further reduction of basis and approximations of the full theory are discussed. To demonstrate the implementation and effect, in Sec.V, we apply the sc-RPA to one-dimensional spinless fermion model, present the computation details, benchmark the results with exact ones, and draw conclusion of the study. Finally, in Sec.VI, we give a summary and discussion.

\end{section}

\begin{section}{EOM of retarded GF and PTA}

In this section, we present the EOM of Bosonic retarded GF and the PTA formalism quite generally. The counterpart for Fermionic retarded GF has been presented in Ref.\cite{Fan1}.

\begin{subsection}{EOM of Bosonic Retarded GF}

Suppose we have a Hamiltonian $H$. We want to calculate the following Boson-type GF (also called commutator GF) of two arbitrary operators $A$ and $B$,
\begin{equation}   \label{Eq1}
   G^{r}[A(t)|B(t^{\prime})] \equiv \frac{1}{i} \theta(t-t^{\prime}) \langle \left[A(t), B(t^{\prime}) \right]\rangle.
\end{equation}
Here, $A(t)=e^{iHt}Ae^{-iHt}$ is the Heisenberg operator and the same for $B(t^{\prime})$. 
$\theta(t-t^{\prime})$ is the step function. $\langle ... \rangle$ denotes the average on the thermal equilibrium state of the Hamiltonian $H$ at temperature $T$. $\left[X,Y\right]=XY-YX$ is the commutator of $X$ and $Y$. We have adopted the unit $\hbar=k_{B}=1$.

The Fourier transformation of $G^{r}[A(t)|B(t^{\prime})]$ is denoted as $G^{r}(A|B)_{\omega}$,
\begin{equation}   \label{Eq2}
   G^{r}(A|B)_{\omega} = \int_{-\infty}^{\infty}  G^{r}[A(t)|B(t^{\prime})] e^{i (\omega + i\eta)(t-t^{\prime}) } d(t-t^{\prime}).
\end{equation}
Here $\eta = 0^{+}$ is an infinitesimal positive number. In terms of Zubarev GF, the EOM reads
\begin{eqnarray}    \label{Eq3}
  \omega   G(A|B)_{\omega} &=& \langle \left[A, B\right] \rangle +  G([A,H]|B)_{\omega}   \nonumber \\
  &=& \langle \left[A, B\right] \rangle -  G(A| [B, H])_{\omega} .
\end{eqnarray}
The two equations are obtained by taking derivative of $t$ and $t^{\prime}$ for $G^{r}[A(t)|B(t^{\prime})] $, respectively. Their equivalence is a consequence of the time-translation invariance of the equilibrium state. 
Once $G(A|B)_{\omega}$ is obtained, one can calculate the average $\langle BA \rangle$
from the spectral theorem,
\begin{equation}   \label{Eq4}
   \langle BA \rangle = \int_{-\infty}^{\infty} d\omega \frac{1}{e^{\beta \omega} -1} \rho_{A,B}(\omega)  + \langle B_{0} A_{0} \rangle.
\end{equation}
Here, the spectral function $\rho_{A,B}(\omega)$ is defined as
\begin{equation}   \label{Eq5}
 \rho_{A,B}(\omega) = \frac{i}{2\pi} \left[ G(A|B)_{\omega+i\eta}  -  G(A|B)_{\omega -i\eta}\right].
\end{equation}
The operator $A_0$ and $B_0$ are the static components of the operators $A$ and $B$, respectively.
Note that any operator $X$ can be written as a sum of its static component $X_0$ and the dynamic one $X_d$,
\begin{eqnarray}   \label{Eq6}
   X &=& X_0 + X_d,        \nonumber \\
   X_0 &=& \sum_{n,m (E_n = E_m)} \langle n|X|m \rangle \,\, | n \rangle \langle m |,  \nonumber \\
   X_d &=& \sum_{n,m (E_n \neq E_m)} \langle n|X|m \rangle \,\, | n\rangle \langle m |.
\end{eqnarray}
Here, $|m \rangle$ and $E_{m}$ are $H$'s eigenstate and eigenenergy, respectively.
The static component fulfills $\left[X_0, H \right] =0$.
For a given $X$, $X_0$ and $X_d$ are in general unknown. In some cases, $X_0 = 0$ for symmetry reasons.

From the definition, we obtain, for the Bosonic GF and spectral function, $G(A|B)_{\omega} = G(A_d|B_d)_{\omega}$ and $\rho_{A,B}(\omega) = \rho_{A_d, B_d}(\omega)$.
The proof of Eqs.(\ref{Eq4}) and (\ref{Eq5}) and some properties of the static component $A_0$ of an operator $A$ are given in Appendix A. Eq.(\ref{Eq4}) shows that only $\langle B_d A_d \rangle = \langle B A \rangle - \langle B_0 A_0 \rangle$ can be calculated directly from the above bosonic GF via the spectral theorem. The static component contribution $\langle B_0 A_0 \rangle$ has to be calculated from other considerations.

It should be noted that for general cases, Eq.(\ref{Eq4}) applies only to the grand canonical ensemble. For the special case where $[H, \hat{N}] = [A, \hat{N}]=[B, \hat{N}] =0$, Eq.(\ref{Eq4}) applies also to the canonical ensemble. Here $\hat{N}$ is the total electron number operator. This is proved in Appendix B. To derive the sc-RPA, we use the basis operator $\{ a_{\alpha}^{\dagger} a_{\beta} \}$ that commute with $\hat{N}$. So the basic sc-RPA equations derived in this paper apply both to the canonical ensemble and to the grand canonical ensemble. However, in the self-consistent calculation of the single-particle and two-particle density matrices in PTA, people sometimes introduce  additional approximations which may be applicable either to the canonical ensemble (such as Rowe's number operator method) or to the grand canonical ensemble (such as the EOM for GF defined for $a_{\alpha}a_{\beta}$ type operators). In those situations, care make be taken to the applicability of the whole theory to certain ensembles.

\end{subsection}

\begin{subsection}{Projective Truncation Approximation}

The PTA formalism is based on approximately solving the GF through truncating the EOM, and on the self-consistent calculation of averages using the spectral theorem. The truncation of EOM is done by
projection method \cite{Fan1,Ma1}.

For a given problem, we first choose a set of operators $ \{ A_1, A_2,..., A_D \}$ as our basis. $D$ is the number of basis operators.
These operators should be the most relevant excitation operators to the problem that we are considering. In practice, due to the lack of the exact knowledge about the studied system and the limitation of computational cost, we usually select certain convenient and simple basis operators. These operators should also be linearly independent. In case that they are linearly dependent, we need additional steps in PTA to remove the redundant basis.

We write down the EOM for the GF $G(A_i|A_j^{\dagger})_{\omega}$ as 
\begin{equation}   \label{Eq7}
  \omega   G(A_i|A_j^{\dagger})_{\omega} = \langle [A_i, A_j^{\dagger}] \rangle +  G([A_i,H]|A_j^{\dagger} )_{\omega}.
\end{equation}
We approximate the operator $[A_i, H]$ as
\begin{equation}   \label{Eq8}
  [A_i, H] \approx \sum_{k} M_{ki} (A_{k})_d.
\end{equation}
The right-hand side of the above equation only involves the dynamic component since $[A_i, H]_0 = 0$. 
The coefficients $M_{ki}$ are determined by projection. This means that on the right-hand side of Eq.(\ref{Eq8}), we neglect the part of $[A_i, H]$ that is orthogonal to the operator basis. To do the projection, we need to define an inner product between two operators $X$ and $Y$. Here, to be consistent with the existing RPA formalism, we use the commutator inner product
\begin{equation}   \label{Eq9}
   (X|Y) = \langle \left[X^{\dagger}, Y \right] \rangle.
\end{equation}
Here the average $\langle ... \rangle$ is defined on the equilibrium state of $H$ at temperature $T$. For this inner product, we have $(X|Y_0)=(X_0|Y)=(X_0|Y_0)=0$ for arbitrary operators $X$ and $Y$.

Note that the definition of inner product $(X|Y)$ is not unique. The anti-commutator or Mori's definition are the frequently used ones \cite{Mori1}. The commutator inner product of Eq.(\ref{Eq9}) is formally compatible with the Boson-type GF used in this work. It does not meet the mathematical requirement $(X|X) > 0$ (for any $X \neq 0$) for an inner product. This does not invalidate the whole formalism of RPA, but does damage the applicability of RPA in the high symmetry phase. See the discussion in the end of this paper.

Projecting Eq.(\ref{Eq8}) to $A_j$ and defining the inner product matrix $\bf{I}$ and the Liouville matrix $\bf{L}$ as
\begin{eqnarray}   \label{Eq10}
   && I_{ij} \equiv (A_i|A_j),  \nonumber \\
   && L_{ij} \equiv (A_i| [A_j, H]),
\end{eqnarray}
we obtain
\begin{equation}   \label{Eq11}
    \bf{L} = \bf{I} \bf{M}.
\end{equation}
Here we have used $(X|Y) =(X|Y_d)$. For a nonsingular matrix ${\bf I}$, we get ${\bf M = I^{-1}L }$.
Using Eqs.(\ref{Eq7}) and (\ref{Eq8}), the GF matrix is formally solved as
\begin{equation}    \label{Eq12}
   G\left(\vec{A}|\vec{A}^{\dagger}\right)_{\omega} \approx - \left[ \omega \bf{1} - \bf{M}^{T} \right]^{-1} \bf{I}^{T}.
\end{equation}
In the above equation, the vector $\vec{A}$ is the column vector $\vec{A} = (A_1, A_2, ..., A_D)^{T}$. 
This is the approximate GF matrix obtained from the truncation of the Heisenberg EOM in Eq.(\ref{Eq8}). The poles of this GF matrix is given by the eigen values of the dynamic matrix ${\bf M}$, and the weights of the GF matrix is determined by the inner product matrix ${\bf I}$ and the eigen vectors of ${\bf M}$. 

Note that the inner product matrix $\bf{I}$ and the Liouville matrix $\bf{L}$ are both hermitian matrices. $\bf{L}$ is negative semi-definite, as proved in Appendix C. In general, ${\bf I}$ is neither positive nor negative semi-definite. In case that $\bf{I}$ is nonsingular, $\bf{M}= \bf{I}^{-1}\bf{L}$ should have real eigenvalues only, since these eigenvalues are the poles of GF in Eq.(\ref{Eq12}). 

Applying the projective truncation to the right-hand side EOM $\omega   G(A_i|A_j^{\dagger})_{\omega} = \langle [A_i, A_j^{\dagger}] \rangle -  G(A_i| [A_j^{\dagger}, H] )_{\omega}$, we can get a similar equation
\begin{equation}    \label{Eq12p}
   G\left(\vec{A}|\vec{A}^{\dagger}\right)_{\omega} \approx - \bf{I}^{T} \left[ \omega \bf{1} - \bf{M}^{\ast} \right]^{-1} .
\end{equation}
This approximate expression of the GF matrix is also due to the truncation of Heisenberg EOM in Eq.(\ref{Eq8}). One can prove that the equivalence of Eqs.(\ref{Eq12}) and (\ref{Eq12p}) amounts to $\bf{L}^{\dagger}=\bf{L}$, which is equivalent to the conservation relation $\partial \langle [A_i^{\dagger}, A_j ] \rangle(t) / \partial t= 0$ for $i,j \in [1, D]$.

Applying the spectral theorem Eqs.(4) and (5) to the GF matrix $G(\vec{A}|\vec{A}^{\dagger})_{\omega}$, we obtain the exact formula
\begin{equation}   \label{Eq13}
   \langle A_{j}^{\dagger}A_{i} \rangle = \int_{-\infty}^{\infty}  d\omega \frac{1}{e^{\beta \omega} -1} \rho_{A_i,A_{j}^{\dagger}}(\omega) + \langle (A_{j})_0^{\dagger} (A_{i})_0 \rangle,
\end{equation}
and
\begin{equation}   \label{Eq14}
 \rho_{A_i,A_{j}^{\dagger}}(\omega) = \frac{i}{2\pi} \left[ G(A_i|A_{j}^{\dagger})_{\omega+i\eta}  -  G(A_i|A_{j}^{\dagger})_{\omega -i\eta}\right].
\end{equation}
If $\bf{I}$ and $\bf{L}$ are expressible in terms of $\langle A_{j}^{\dagger}A_{i} \rangle$ ($i,j = 1,2,..., D$), and if $\{ \langle (A_{j})_0^{\dagger} (A_{i})_0 \rangle \}$ are known, Eqs.(\ref{Eq10})-(\ref{Eq14}) form a closed set of equations about the averages $\{ \langle A_{j}^{\dagger}A_{i} \rangle \}$. Solving these equations, one obtains the approximate values of these averages as well as $\bf{M}$ and $G (\vec{A}|\vec{A}^{\dagger} )_{\omega}$.

One can further analyse the eigen-excitation operator $O_{\nu}$ associated to the excitation energy $\lambda_{\nu}$. Suppose we find a matrix $\bf{U}$ such that
\begin{equation}   \label{Eq15}
   \bf{U}^{-1} {\bf M} {\bf U} = {\bf \Lambda} \equiv \text{diag}(\lambda_1, \lambda_2, ..., \lambda_D).
\end{equation}
${\bf \Lambda}$ is a real diagonal matrix \cite{Note1}. 
By Eq.(\ref{Eq11}), this equation can be transformed into the generalized eigen-value problem
\begin{equation}    \label{Eq17p}
   {\bf L U = I U \Lambda }.
\end{equation}
If we define the following combinations of the basis operators
\begin{equation}   \label{Eq16}
   O_{\nu} \equiv \sum_{i=1}^{D} U_{i \nu} (A_{i})_d,
\end{equation}
From Eqs.(\ref{Eq8}) and (\ref{Eq15}), we obtain
\begin{equation}    \label{Eq17}
  [O_{\nu}, H] \approx \lambda_{\nu} O_{\nu}.
\end{equation}
That shows that $O_{\nu}$ is the approximate eigen excitation operator with the excitation energy $\lambda_{\nu}$.
This equation is nothing but the truncation approximation Eq.(\ref{Eq8}) on the diagonal operator basis.

Here, the excitation energies $\{ \lambda_{\nu} \}$ are the eigenvalue of $\bf{M}$ and the poles of GF matrix $G(\vec{A}|\vec{A}^{\dagger})_{\omega}$.
$O_{\nu}$ with $\lambda_\nu < 0$ corresponds to the {\it excitation} operator, i.e., it raises the energy of a state by $-\lambda_\nu$.

Although $\{ O_{\nu} \}$ are the approximate eigen excitation operators, they are exactly orthogonal to each other. First, it is easy to prove that $(O_{\nu}|O_{\nu^{\prime}} ) = 0$ for $\lambda_{\nu} \neq \lambda_{\nu^{\prime}}$. Then, for degenerate eigen-excitation operators, one can diagonalize their inner product sub-matrix to make them orthogonal to each other. We can then write
\begin{equation}   \label{Eq18}  
  \left( O_{\nu} | O_{\nu^{\prime}} \right) = \left( O_{\nu} | O_{\nu} \right)\delta_{\nu, \nu^{\prime}}.
\end{equation}
From Eqs.(\ref{Eq12}), (\ref{Eq13}), (\ref{Eq14}), and (\ref{Eq18}), we obtain
\begin{eqnarray} 
   && \rho_{O_{\nu^{\prime}}, O_{\nu}^{\dagger}}(\omega) \approx - \left( O_{\nu} | O_{\nu} \right) \delta(\omega - \lambda_{\nu}) \delta_{\nu, \nu^{\prime}},  \label{Eq19} \\ 
   && \langle O_{\nu}^{\dagger} O_{\nu^{\prime}} \rangle \approx - \frac{\left( O_{\nu} | O_{\nu} \right)}{e^{\beta \lambda_{\nu}}-1 }\delta_{\nu, \nu^{\prime}} .  \label{Eq20}
\end{eqnarray}

The normalization of eigen-excitation operators corresponds to determining the free factor of each column of ${\bf U}$. By requiring the normalization condition $(O_{\nu}|O_{\nu}) = \pm 1$ ($\nu = 1,2,..., D$) and using Eq.(\ref{Eq16}), we obtain the generalized orthonormal condition for ${\bf U}$ as ${\bf U}^{\dagger} {\bf I U} = \text{diag}\{ \pm 1, \pm 1, ..., \pm 1 \}$. Now the eigen excitation operator $O_{\nu}$ has been fixed up to an arbitrary phase factor $e^{i \theta_{\nu}}$. In summary, by selecting appropriate ${\bf U}$, one can simultaneously diagonalize ${\bf I}$, ${\bf L}$, and ${\bf M}$,
\begin{eqnarray}   \label{Eq20p}
&& {\bf U}^{\dagger} {\bf I U} =  \text{diag}\{ \pm 1, \, \pm 1, \, ..., \, \pm 1 \} , \nonumber \\
&& {\bf U}^{\dagger} {\bf L U} = \text{diag}\{ l_1, \, l_2, \, ...,\,  l_D \},  \nonumber \\
&& {\bf U}^{-1} {\bf M U} = \bf{\Lambda} \equiv \text{diag}\{ \lambda_{1}, \, \lambda_2, \, ..., \, \lambda_{D} \}.
\end{eqnarray}
With this transformation, the basis operators are transformed into orthonormal eigen-excitation operators $\{ O_{\nu} \}$ and Eqs.(\ref{Eq18}-\ref{Eq20}) are simplified.

Using matrices ${\bf I}$ and ${\bf M}$, the averages in Eq.(\ref{Eq13}) is expressed as
\begin{equation}    \label{Eq20.1}
   \langle A_{j}^{\dagger}A_{i} \rangle \approx - \sum_{k=1}^{D}I_{jk} \left[ \left( e^{\beta {\bf M}} - \bf{1} \right)^{-1} \right]_{ki} + \langle (A_{j})_0^{\dagger} (A_{i})_0 \rangle.
\end{equation}
In terms of the matrix ${\bf C}_{ji} \equiv \langle A_j^{\dagger} A_i \rangle$ and $[{\bf C}_0]_{ji} \equiv \langle (A_{j})_0^{\dagger} A_{i 0} \rangle$, Eq.(\ref{Eq20.1}) is written into a concise form,
\begin{equation}    \label{Eq20.2}
   {\bf C } \approx - {\bf I} \left( e^{\beta {\bf M}} - \bf{1} \right)^{-1} + {\bf C}_0,
\end{equation}
which forms a closed set of equations with ${\bf L = L(C)} $ and ${\bf I = I(C)}$.
Note that the approximations in the above equations Eqs.(\ref{Eq17}), (\ref{Eq19}), (\ref{Eq20}), (\ref{Eq20.1}), and (\ref{Eq20.2}) are entirely due to the truncation approximation made in Eq.(\ref{Eq8}). No additional approximations are introduced. Therefore, in the limit where the operator basis becomes complete, these equations will become exact.

The actual PTA calculation can thus be divided into two successive steps, the thermodynamic step and the dynamical step. In the thermodynamic step, one solves Eq.(\ref{Eq20.2}) to obtain ${\bf M}$ and ${\bf I}$. Then, in the dynamical step, one computes the GFs Eq.(\ref{Eq12p}) and spectral functions Eq.({\ref{Eq14}}).

In the above derivation, we assume that the inner product matrix ${\bf I}$ is non-singular. If ${\bf I}$ has zero eigen-values, the matrix $\bf{M}$ is not unique. This occurs when the basis operators are linearly dependent, or certain symmetry between the basis operators exist. In either case, one can solve Eq.(\ref{Eq17p}) by using the standard technique of removing the redundant basis operators. Such method has been frequently used in the energy band calculations with non-orthogonal basis.
It amounts to removing the zero-length operators in the ${\bf I}$-diagonalized basis, rebuilding the matrix $\bf{I}$, $\bf{L}$, and ${\bf M}$ on the remaining operators of finite length, and carrying out PTA on the reduced operator subspace. Our application of this method in the study of classical one-dimensional $\phi^4$ lattice model shows that this technique can substantially stabilize the self-consistent calculation \cite{Jia1}.

\end{subsection}

\end{section}

\begin{section}{RPA on the basis $\{ a_{\alpha}^{\dagger}a_{\beta} \}$ ($\alpha \neq \beta$)}

The above PTA formalism is quite general. Below, we will take a special operator basis,  $\{ a_{\alpha}^{\dagger}a_{\beta} \}$ ($\alpha \neq \beta$). We will show that under this basis, the PTA produces sc-RPA.  

\begin{subsection}{Formalism on Arbitrary Orbitals}

Suppose we define the fermionic single particle creation operators $ \{ a_{\alpha}^{\dagger} \}$ and the annihilation operators $\{ a_\alpha \}$ on a set of given orthonormal single-particle orbitals $\{ |\phi_\alpha \rangle \}$.
Here $\alpha = 1,2,..., L$ includes all the single particle indices such as spin, orbital, momentum, {\it etc.}
$a_{\alpha}^{\dagger}$ and $a_\alpha$ obey the usual anti-commutator relations $\{a_{\alpha}, a_{\beta}^{\dagger} \} = \delta_{\alpha, \beta}$. The Hamiltonian that we are studying $H[\{a_{\alpha}^{\dagger}, a_{\alpha} \}]$ is written in terms of these operators. For the moment, we do not need to specify these orbitals since the formalism developed in this subsection applies for general orbitals.   

We now assign the orbitals \{$\alpha$\} an order such that the single particle occupation number $n_{\alpha} \equiv \langle a_{\alpha}^{\dagger} a_{\alpha} \rangle$ decreases with increasing $\alpha$. That is, $n_{\alpha} \geq n_{\beta}$ for $\alpha < \beta$. The most general basis considered in this work is $\mathcal{B}_1 \equiv \{ a_{\alpha}^{\dag} a_{\beta} \}$ ($\alpha \neq \beta$).
In case that there are occupation degeneracies, the operators $a_{\alpha}^{\dagger}a_{\beta}$ ($\alpha \neq \beta$) with $n_{\alpha} = n_{\beta}$ will be discarded from the basis. The dimension of the basis is $D = L^{2} - \sum_{i} d_{i}^{2}$, where $d_i$ is the number of orbitals with the same occupation $n_i$. In this work, we only consider $d_i =1$ case, i.e., the case with no occupancy degeneracy. The dimension of the basis set is $D = L(L-1)$. 
 
We denote $A_{i} = a_{\alpha}^{\dagger} a_{\beta}$ and $B_{i} = a_{\beta}^{\dagger} a_{\alpha} = A_{i}^{\dagger}$ ($\alpha > \beta$). Below we use $i$ to represent a pair of Greek index $i=(\alpha > \beta)$. The basis operators are ordered as 
$\mathcal{B}_{1} = \{ A_{1}, A_{2}, ..., A_{D/2}, B_1, B_2, ..., B_{D/2} \}$. Sometimes, for symmetry reasons, $\{A_i\}$ and $\{B_i\}$ belong to two subsets (see the formula for the spinless fermion model below). For the moment, we will consider the most general situation. With this sequence of basis operators, the inner product matrix $\bf{I}$ and Liouville matrix $\bf{L}$ become $2 \times 2$ block matrices,
\begin{equation}    \label{Eq21}
  {\bf I} = \left(
\begin{array}{cc}
I_{11} &  I_{12}  \\
I_{21} &  I_{22}  \\
\end{array}
\right),
\end{equation}
and
\begin{equation}    \label{Eq22}
  {\bf L} = \left(
\begin{array}{cc}
L_{11} &  L_{12}  \\
L_{21} &  L_{22}  \\
\end{array}
\right).
\end{equation}
Denoting $i = (\alpha > \beta)$ and $j = (\alpha^{\prime} > \beta^{\prime})$, straightforward calculation gives
\begin{eqnarray}  
   (I_{11})_{ij} & \equiv & (A_i|A_j) = \delta_{\alpha \alpha^{\prime}} \langle a_{\beta}^{\dagger} a_{\beta^{\prime}} \rangle - \delta_{\beta \beta^{\prime}} \langle a_{\alpha^{\prime}}^{\dagger} a_{\alpha} \rangle, \label{Eq23} \\
   (I_{12})_{ij} & \equiv & (A_i|B_j) = \delta_{\alpha \beta^{\prime}} \langle a_{\beta}^{\dagger} a_{\alpha^{\prime}} \rangle - \delta_{\beta \alpha^{\prime}} \langle a_{\beta^{\prime}}^{\dagger} a_{\alpha} \rangle. \label{Eq24}
\end{eqnarray}
Here $i,j = 1,2,..., D/2$. 
Due to the Hermiticity of $\bf{I}$, $\bf{I}$ has the following structure
\begin{equation}    \label{Eq25}
  {\bf I} = \left(
\begin{array}{cc}
I_{11} &  I_{12}  \\
-I_{12}^{\ast} &  -I_{11}^{\ast}  \\
\end{array}
\right),
\end{equation}
with additional constraints $I_{11}^{\dagger} = I_{11}$ and $I^{T}_{12}=-I_{12}$. ${\bf I}$ has both positive and negative eigenvalues.

Similarly, one can prove without explicitly writing down the expression of ${\bf L}$ that 
\begin{equation}    \label{Eq26}
  {\bf L} = \left(
\begin{array}{cc}
L_{11} &  L_{12}  \\
L_{12}^{\ast} &  L_{11}^{\ast}  \\
\end{array}
\right).
\end{equation}
Here, $(L_{11})_{ij} = (A_i|[A_j, H])$ and  $(L_{12})_{ij} = (A_i|[B_j, H])$.
There are additional constraints $L_{11}^{\dagger} = L_{11}$ and $L^{T}_{12}= L_{12}$. Explicit expression for $(L_{11})_{ij}$ and $(L_{12})_{ij}$ will be given in Sec.III.C after the Hamiltonian $H$ is specified.
The above structure of $\bf{I}$ and $\bf{L}$ can be summarized as $\bf{J}\bf{I}\bf{J} = -\bf{I}^{\ast}$ and $\bf{J}\bf{L}\bf{J} = \bf{L}^{\ast}$,
with 
\begin{equation}    \label{Eq27}
  {\bf J} = \left(
\begin{array}{cc}
\bf{0} & \bf{1}  \\
\bf{1} & \bf{0}  \\
\end{array}
\right).
\end{equation}

In the generalized eigen-value problem Eq.(\ref{Eq17p}), employing the structure Eqs.(\ref{Eq25}) and (\ref{Eq26}), we can require that $\bf{U}$ and $\bf{\Lambda}$ fulfil similar structure, i.e., $\bf{J}\bf{U}\bf{J}=\bf{U}^{\ast}$ and $\bf{J}\bf{\Lambda}\bf{J}= - \bf{\Lambda}^{\ast}$. We therefore can write
\begin{equation}    \label{Eq29}
  {\bf U} = \left(
\begin{array}{cc}
U_{11} &  U_{12}  \\
U_{12}^{\ast} &  U_{11}^{\ast}  \\
\end{array}
\right),
\end{equation}
and
\begin{equation}    \label{Eq30}
  {\bf \Lambda} = \left(
\begin{array}{cc}
\Lambda_{11} &  \bf{0}  \\
\bf{0} &  - \Lambda_{11}^{\ast}  \\
\end{array}
\right),
\end{equation}
Since $\Lambda_{11}$ is real, the eigenvalues of $\bf{M}$ appear in plus-minus pairs.
To be consistent with Rowe's formalism, below we assign $(\Lambda_{11})_{\nu \nu} \equiv \lambda_{1\nu} < 0$ and $(\Lambda_{22})_{\nu \nu} \equiv  \lambda_{2\nu} = - \lambda_{1\nu} > 0$. This means that $O_{1 \nu}$ operator is the {\it excitation} operator of the energy spectrum.
At zero temperature, it corresponds to the operator $Q_{\nu}^{\dagger}$ of Rowe. That is, it fulfils $O_{1\nu}|0\rangle \approx |\nu \rangle$, where $|0 \rangle$ and $| \nu \rangle$ are the ground state and the excited state of $H$, respectively..

With the above form of ${\bf U}$ matrix, Eq.(\ref{Eq16}) becomes
\begin{eqnarray}   \label{Eq31}
   && O_{1\nu} = \sum_{i=1}^{D/2} \left[ (U_{11})_{i \nu} (A_{i})_d + (U_{12}^{\ast})_{i \nu} (B_{i})_d \right],  \nonumber \\
   && O_{2\nu} = \sum_{i=1}^{D/2} \left[ (U_{12})_{i \nu} (A_{i})_d + (U_{11}^{\ast})_{i \nu} (B_{i})_d \right] = O_{1\nu}^{\dag},  \nonumber \\
  && 
\end{eqnarray}
with $1 \leq \nu \leq D/2$. They fulfil
$[O_{1\nu}, H] \approx \lambda_{1\nu} O_{1\nu}$ and $[O_{2\nu}, H] \approx \lambda_{2\nu} O_{2\nu}$, with $\lambda_{2 \nu} = - \lambda_{1\nu}$.    

These eigen-excitation operators are naturally orthogonal. Further requiring that they are normalized, i.e.,
\begin{eqnarray}   \label{Eq31.5}
 && (O_{1\nu}|O_{1\nu^{\prime}}) = \delta_{\nu, \nu^{\prime}},   \nonumber \\
 && (O_{2\nu}|O_{2\nu^{\prime}}) = -\delta_{\nu\nu^{\prime}},  \nonumber \\
 && (O_{1\nu}|O_{2\nu^{\prime}}) = 0,
\end{eqnarray}
we obtain the average values for $\langle O_{1\nu}^{\dagger} O_{1\nu} \rangle$ 
and $\langle O_{2\nu}^{\dagger} O_{2\nu} \rangle$ ($1 \leq \nu \leq D/2$) from Eq.(\ref{Eq20}) as
\begin{eqnarray}
   && \langle O_{1\nu}^{\dagger} O_{1\nu^{\prime}} \rangle \approx \frac{-1}{e^{\beta \lambda_{1 \nu}}-1 } \delta_{\nu \nu^{\prime}},     \label{Eq32}  \\
   && \langle O_{2\nu}^{\dagger} O_{2\nu^{\prime}} \rangle \approx \frac{1}{e^{-\beta \lambda_{1 \nu}} -1 } \delta_{\nu \nu^{\prime}},     \label{Eq33}   \\
      && \langle O_{1\nu}^{\dagger} O_{2\nu^{\prime}} \rangle = \langle O_{2\nu}^{\dagger} O_{1\nu^{\prime}} \rangle \approx  0.    \label{Eq34}  
\end{eqnarray}
At $T=0$, these equations recover the corresponding ones of Rowe \cite{Rowe1}.
Similarly, the spectral function of eigen-excitation operator is obtained from Eq.(\ref{Eq19}) as
\begin{eqnarray}
   && \rho_{O_{1\nu}, O_{1\nu^{\prime}}^{\dagger} }(\omega) \approx - \delta_{\nu \nu^{\prime}} \delta(\omega - \lambda_{1\nu}),        \label{Eq35}  \\
   && \rho_{ O_{2\nu}, O_{2\nu^{\prime}}^{\dagger} }(\omega) \approx  \delta_{\nu \nu^{\prime}} \delta(\omega + \lambda_{1\nu}),      \label{Eq36}  \\
      && \rho_{ O_{1\nu}, O_{2\nu^{\prime}}^{\dagger} }(\omega) = \rho_{ O_{2\nu}, O_{1\nu^{\prime}}^{\dagger} }(\omega)\approx  0.    \label{Eq37}  
\end{eqnarray}
These equations show that the statistical and dynamical properties of the eigen-operators $O_{1\nu}$ ($O_{2\nu}$) are similar to those of a canonical boson operator $a^{\dagger}_{\nu}$ ($a_{\nu}$), with occupation energy $-\lambda_{1\nu}$. This is the foundation of the frequently used quasi-boson approximation in the Rowe formalism.

The orthonormalization of $O_{1\nu}$ and $O_{2\nu}$ Eq.(\ref{Eq31.5}) implies that $\bf{U}$ has the general orthonormal properties. Inserting Eq.(\ref{Eq31}) into Eq.(\ref{Eq31.5}) and employing the structures of $\bf{U}$ and $\bf{I}$, we obtain the constraints to $\bf{U}$ matrix as 
\begin{equation}    \label{Eq40}
  {\bf U}^{\dagger} \bf{I} \bf{U} = \left(
\begin{array}{cc}
\bf{1} &  \bf{0}  \\
\bf{0} &  - \bf{1}  \\
\end{array}
\right).
\end{equation}
Note that Eq.(\ref{Eq40}) is different from the generalized orthonormal relation $\bf{U}^{\dagger} \bf{I} \bf{U}=\bf{1}$ for the generalized eigen-value problem $\bf{L}\bf{U}=\bf{I}\bf{U}\bf{\Lambda}$ with a positive-definite inner product matrix ${\bf I}$. The latter appears in the PTA for Fermionic type basis \cite{Fan1}.

The inverse of $\bf{U}$ reads
\begin{eqnarray}
    {\bf U}^{-1} &=& \left(
\begin{array}{cc}
\bf{1} &  \bf{0}  \\
\bf{0} &  - \bf{1}  \\
\end{array}
\right) \bf{U}^{\dagger} \bf{I}  \nonumber \\
 &= & \left(
 \begin{array}{cc}
U_{11}^{\dagger}I_{11} - U_{12}^{T} I_{12}^{\ast} &  U_{11}^{\dagger}I_{12} - U_{12}^{T} I_{11}^{\ast}  \\
U_{11}^{T}I_{12}^{\ast} - U_{12}^{\dagger} I_{11} &  U_{11}^{T}I_{11}^{\ast} - U_{12}^{\dagger} I_{12}  \\
\end{array}
\right).      \label{Eq41}  
\end{eqnarray}
The closure relation of $O_{1\nu}$ and $O_{2 \nu}$ is obtained by left-multiplying $\bf U$ to both sides of Eq.(\ref{Eq41}). We obtain
\begin{eqnarray}
&& \left(U_{11} U_{11}^{\dagger} - U_{12} U_{12}^{\dagger} \right) I_{11} -  \left(U_{11} U_{12}^{T} - U_{12} U_{11}^{T} \right) I_{12}^{\ast} = \bf{1},  \nonumber \\
&&  \left(U_{11} U_{11}^{\dagger} - U_{12} U_{12}^{\dagger} \right) I_{12} -  \left(U_{11} U_{12}^{T} - U_{12} U_{11}^{T} \right) I_{11}^{\ast} = \bf{0}.    \nonumber \\
&&     \label{Eq42}  
\end{eqnarray}
Using $\bf{J}\bf{U}^{-1} \bf{J}=(\bf{U}^{-1})^{\ast}$,  we express the dynamical component of the basis operators in terms of the eigen-excitation operators as
\begin{eqnarray}  \label{Eq43}
   && (A_{i})_d = \sum_{\nu=1}^{D/2} \left\{ [(U^{-1})_{11} ]_{\nu i} O_{1\nu} + [(U^{-1})_{12}^{\ast} ]_{\nu i} O_{2\nu}\right\},  \nonumber \\
   && (B_{i})_d = \sum_{\nu=1}^{D/2} \left\{ [(U^{-1})_{12} ]_{\nu i} O_{1\nu} + [(U^{-1})_{11}^{\ast} ]_{\nu i} O_{2\nu}\right\}.  \nonumber \\
   &&
\end{eqnarray}

The spectral functions of GF matrix $G(A_{i}|A_{j}^{\dagger})$ are then obtained from the above expression and Eqs.(\ref{Eq35})-(\ref{Eq37}) as
\begin{eqnarray}
   && \rho_{A_i, A_j^{\dagger}}(\omega)   =  \rho_{(A_{i})_d, (A_{j})_d^{\dagger}}(\omega)\nonumber \\
   &  \approx & - \sum_{\nu=1}^{D/2} [(U^{-1})_{11}]_{\nu i} [(U^{-1})_{11}]^{\ast}_{\nu j} \delta(\omega - \lambda_{1 \nu})   \nonumber \\
   &&  + \sum_{\nu=1}^{D/2} [(U^{-1})_{12}]^{\ast}_{\nu i} [(U^{-1})_{12}]_{\nu j} \delta(\omega + \lambda_{1 \nu}) ,   \label{Eq44}
\end{eqnarray}
and
\begin{eqnarray}
   && \rho_{A_i, B_j^{\dagger}}(\omega) =\rho_{(A_{i})_d, (B_{j})_d^{\dagger}}(\omega)       \nonumber \\
   &  \approx & - \sum_{\nu=1}^{D/2} [(U^{-1})_{11}]_{\nu i} [(U^{-1})_{12}]^{\ast}_{\nu j} \delta(\omega - \lambda_{1 \nu})   \nonumber \\
   &&  + \sum_{\nu=1}^{D/2} [(U^{-1})_{12}]^{\ast}_{\nu i} [(U^{-1})_{11}]_{\nu j} \delta(\omega + \lambda_{1 \nu}) .  \label{Eq45}
\end{eqnarray}
The other spectral functions are related to the above two by  $\rho_{B_i, B_j^{\dagger}}(\omega) = -\rho^{\ast}_{A_i, A_j^{\dagger}}(-\omega)$ and $\rho_{B_i, A_j^{\dagger}}(\omega) = -\rho^{\ast}_{A_i, B_j^{\dagger}}(-\omega)$.
These spectral functions are related to the experimental observables such as dielectric functions, dynamical magnetic susceptibilities, optical conductivity, optical absorption, {\it etc.}, depending on the choice of operators.

The averages of the type $\langle (A_{i})_d^{\dagger} (A_{j})_d\rangle$ can be obtained from Eq.(\ref{Eq44}) as
\begin{eqnarray}
   && \langle (A_{i})_d^{\dagger} (A_{j})_d \rangle    \nonumber \\
   &  \approx & - \sum_{\nu=1}^{D/2} [(U^{-1})_{11}]_{\nu i}^{\ast} [(U^{-1})_{11}]_{\nu j} \frac{1}{e^{\beta \lambda_{1\nu}}-1}   \nonumber \\
   &&  + \sum_{\nu=1}^{D/2} [(U^{-1})_{12}]_{\nu i} [(U^{-1})_{12}]^{\ast}_{\nu j} \frac{1}{e^{-\beta \lambda_{1\nu}}-1} .   \label{Eq46}
\end{eqnarray}
$\langle (B_{j})_d^{\dagger} (A_{i})_d\rangle$, $\langle (B_{j})_d^{\dagger} (B_{i})_d\rangle$, and  $\langle (A_{j})_d^{\dagger} (B_{i})_d\rangle$ can be written down similarly. 
Similar to the general formalism presented in Sec.II B, the approximations in the equations Eqs.(\ref{Eq32})-(\ref{Eq37}) and (\ref{Eq44})-(\ref{Eq46}) are entirely due to the truncation approximation made in Eq.(\ref{Eq8}) and will be come exact in the complete basis limit. 

Now we consider how to self-consistently determine the matrix $\bf{I}$ and $\bf{L}$. For a Hamiltonian with one-body term and two-body interaction, the matrix elements of $\bf{I}$ and $\bf{L}$ contain the averages of the type $\langle a_{\alpha}^{\dagger} a_{\beta} \rangle$ and $\langle a_{\alpha}^{\dagger}a_{\beta} a_{\gamma}^{\dagger}a_{\delta} \rangle$, that is, the one- and two-particle density matrices. In the self-consistent version of RPA, these quantities are to be calculated self-consistently from the GFs and spectral theorem. 
However, there is one problem to be solved in this process. The spectral theorem Eq.(\ref{Eq13}) or Eq.(\ref{Eq46}) can only produce the dynamical contribution of the type $\langle (A_{i})_d^{\dagger}(A_{j})_d \rangle$ and the static contribution $\langle (A_{i})_0^{\dagger}(A_{j})_0 \rangle$ cannot be produced. In some cases, $(A_{i})_0=0$ due to symmetry reasons (see the one-dimensional spinless fermion model studied below). But in general $(A_{i})_0 \neq 0$.

To solve this problem, the easiest way is to introduce an approximation: We simply ignore the contribution from the static component in the spectral theorem, i.e.,
\begin{eqnarray}
   \langle (A_{i})_0^{\dagger} (A_{j})_0 \rangle \approx 0  \label{Eq47}
\end{eqnarray}
for all the basis operators in our calculation $i,j \in [1,D]$. 
In order to minimize the error introduced by this approximation, we will use the natural orbitals to define our single-particle annihilation and creation operators. The natural orbital for arbitrary temperature is defined as the orbitals on which the one-particle density is diagonal,
\begin{equation}
   \langle a_{\alpha}^{\dagger} a_{\beta}\rangle = \langle a_{\alpha}^{\dagger} a_{\alpha} \rangle \delta_{\alpha \beta}.   \label{Eq48}
\end{equation}
On the natural orbitals, we have $\langle (A_{i})_0 \rangle = \langle A_i \rangle = 0$. The approximation Eq.(\ref{Eq47}) amounts to ignoring the statistical correlation between the static component of basis operators.
In particular, if the ground state has no degeneracy, we have $(A_{i})_0 |0 \rangle = 0$ for $A_i$ defined on natural orbitals. Then Eq.(\ref{Eq47}) becomes exact.
In Rowe's formalism \cite{Rowe1}, the killing condition $ Q_{\nu} |0 \rangle = 0$ and the requirement $|\nu \rangle = Q_{\nu}^{\dagger} | 0 \rangle$ ($|0 \rangle$ and $| \nu \rangle$ are the ground state and excited state, respectively) shows that the same approximation has been used implicitly.

At finite temperature, or at $T=0$ with ground state degeneracy, Eq.(\ref{Eq47}) is no longer exact. But we expect that $\langle (A_{i})_0^{\dagger} (A_{j})_0 \rangle$ is smaller on the natural orbitals than on other orbitals. The sc-RPA equations are also simplified significantly on natural orbitals. Therefore, in the discussion below, we will only use natural orbital. The issue of self-consistent determination of the natural orbitals will be discussed in Sec.III.D.

\end{subsection}

\begin{subsection}{Formalism on Natural Orbitals}

In this part as well as in the rest of the paper, we will use the natural orbital in the PTA formalisms. The PTA formalism developed above will get simplified under the natural orbital. We will also make contact with Rowe's formalism. 

In general, the natural orbitals are unknown {\it apriori}. We need to find them by iterative calculations. For a given set of initial orbitals (usually not the true natural orbitals), we carry out PTA calculation using the formula presented below, which assumes that the orbitals are natural orbitals. Then we evaluate the one-particle density matrix ${\bold \rho}$ defined as $\rho_{\alpha \beta} \equiv \langle a_{\alpha}^{\dagger} a_{\beta}\rangle$ ($1 \leq \alpha, \beta \leq L$). we diagonalize $\bf{\rho}$ to produce the orbital on which $\bf{\rho}$ is diagonal. This set of orbitals is closer to the true natural orbital. We update the orbital and write down the Hamiltonian on the new orbitals. For this new Hamiltonian PTA calculation is carried out again. This goes on until the orbitals converge. This is why we have to evaluate off-diagonal densities like $\langle a_{\alpha}^{\dagger} a_{\beta}\rangle$ ($\alpha \neq \beta$) even if our formalism below assumes that the orbitals are natural orbitals. Details of the orbital update is given in Sec.III.D.

Under the natural orbital, some of the equations derived in previous subsections can be simplified.
We have for the inner product matrix $\bf{I}$
\begin{equation}
  {\bf I} = \left(
\begin{array}{cc}
I_{11} &  \bf{0}  \\
\bf{0} &  - I_{11}  \\
\end{array}
\right),   \label{Eq49}  
\end{equation}
and $(I_{11})_{ij} = (n_{\beta} - n_{\alpha} ) \delta_{ij} $ is a real number. Here $i$ and $j$ are used to denote the ordered pair of orbitals, i.e., $i \equiv (\alpha > \beta)$ and $j \equiv (\alpha^{\prime} > \beta^{\prime})$. We have $\delta_{ij} = \delta_{\alpha \alpha^{\prime}} \delta_{\beta \beta^{\prime}}$. $n_{\alpha} \equiv \langle a_{\alpha}^{\dagger} a_{\alpha} \rangle$ is the average electron occupation of $\alpha$ orbital, and similarly for $n_{\beta}$. Since we have assumed that $n_{\alpha}$ decreases with increasing $\alpha$, $I_{11}$ has positive diagonal. 

The generalized eigen-value problem now has the following form,
\begin{eqnarray}
&&   \left(
\begin{array}{cc}
L_{11} &  L_{12}  \\
L_{12}^{\ast} &  L_{11}^{\ast}  \\
\end{array}
\right)   \left(
\begin{array}{cc}
U_{11} &  U_{12}  \\
U_{12}^{\ast} &  U_{11}^{\ast}  \\
\end{array}
\right)   \nonumber \\
 &=&   \left(
\begin{array}{cc}
I_{11} &  \bf{0}  \\
\bf{0} &  -I_{11}  \\
\end{array}
\right) \left(
\begin{array}{cc}
U_{11} &  U_{12}  \\
U_{12}^{\ast} &  U_{11}^{\ast}  \\
\end{array}
\right)\left(
\begin{array}{cc}
\Lambda_{11} &  \bf{0}  \\
\bf{0} &  - \Lambda_{11}  \\
\end{array}
\right).     \label{Eq50}  
\end{eqnarray}
The generalized orthonormal and closure conditions of $\bf{U}$ becomes, respectively,
\begin{eqnarray}
 && U_{11}^{\dagger}I_{11}U_{11}  - U_{12}^{T}I_{11}U_{12}^{\ast} = \bf{1},   \nonumber \\
 && U_{11}^{\dagger}I_{11}U_{12}  - U_{12}^{T}I_{11}U_{11}^{\ast} = \bf{0}.      \label{Eq51}  
\end{eqnarray}
and 
\begin{eqnarray}
&& \left(U_{11} U_{11}^{\dagger} - U_{12} U_{12}^{\dagger} \right) I_{11} = \bf{1},  \nonumber \\
&& U_{11} U_{12}^{T} - U_{12} U_{11}^{T}  = \bf{0}.      \label{Eq52} 
\end{eqnarray}
$\bf{U}^{-1}$ in Eq.(\ref{Eq41}) is simplified into
\begin{equation}
    {\bf U}^{-1} =  \left(
 \begin{array}{cc}
U_{11}^{\dagger}I_{11}   & - U_{12}^{T} I_{11}  \\
- U_{12}^{\dagger} I_{11} &  U_{11}^{T}I_{11}  \\
\end{array}
\right).    \label{Eq53} 
\end{equation}
Eq.(\ref{Eq43}) is simplified into
\begin{equation}
   (A_{i})_d = (I_{11})_{ii} \sum_{\nu=1}^{D/2} \left[ (U_{11})^{\ast}_{i \nu} O_{1\nu} - (U_{12})^{\ast}_{i \nu} O_{2\nu}\right].     \label{Eq54} 
\end{equation}
The expression for $\langle (A_{i})_d^{\dagger} (A_{j})_d \rangle$ in Eq.(\ref{Eq46}) is simplified in the case of natural orbital into
\begin{eqnarray}
&&   \langle A_{i}^{\dagger} A_{j} \rangle \approx \langle (A_{i})_d^{\dagger} (A_{j})_d \rangle  \nonumber \\
&\approx &  - (I_{11})_{ii}(I_{11})_{jj} \sum_{\nu=1}^{D/2} \left[ (U_{11})_{i\nu}(U_{11})_{j\nu}^{\ast} \frac{1}{e^{\beta \lambda_{1\nu}} -1}\right]  \nonumber \\
&& + (I_{11})_{ii}(I_{11})_{jj} \sum_{\nu=1}^{D/2} \left[ (U_{12})_{i\nu}(U_{12})_{j\nu}^{\ast}  \frac{1}{e^{-\beta \lambda_{1\nu}} -1} \right].      \nonumber \\
&&     \label{Eq55} 
\end{eqnarray}

The spectral function $\rho_{A_i, A_j^{\dagger}}(\omega)$ in Eq.(\ref{Eq44}) now becomes
\begin{eqnarray}
&& \rho_{A_i, A_{j}^{\dagger}}(\omega) \nonumber \\
 & \approx & - (I_{11})_{ii}(I_{11})_{jj} \sum_{\nu=1}^{D/2} \left[(U_{11})^{\ast}_{i\nu}(U_{11})_{j\nu}  \delta(\omega -\lambda_{1\nu}) \right] \nonumber \\
  &&  + (I_{11})_{ii}(I_{11})_{jj} \sum_{\nu=1}^{D/2} \left[(U_{12})^{\ast}_{i\nu}(U_{12})_{j\nu}  \delta(\omega + \lambda_{1\nu}) \right]. \nonumber \\
  &&    \label{Eq56} 
\end{eqnarray}
Eq.(\ref{Eq45}) is simplified into
\begin{eqnarray}   \label{Eq57}
&& \rho_{A_i, B_{j}^{\dagger}}(\omega) \nonumber \\
 & \approx & (I_{11})_{ii}(I_{11})_{jj} \sum_{\nu=1}^{D/2} \left[(U_{11})^{\ast}_{i\nu}(U_{12})_{j\nu}^{\ast}  \delta(\omega -\lambda_{1\nu}) \right] \nonumber \\
  &&  - (I_{11})_{ii}(I_{11})_{jj} \sum_{\nu=1}^{D/2} \left[(U_{12})^{\ast}_{i\nu}(U_{11})_{j\nu}^{\ast}  \delta(\omega + \lambda_{1\nu}) \right]. \nonumber \\
  &&
\end{eqnarray}
Note that Eqs.(\ref{Eq55}) - (\ref{Eq57}) contain not only the error of truncation introduced in Eq.(\ref{Eq8}), but also the error introduced in Eq.(\ref{Eq47}), i.e., from neglecting the contribution of the static component of basis operators.

\end{subsection}

\begin{subsection}{{\bf L} Matrix}

The Hamiltonian of the studied system enters the whole theory only through the $\bf{L}$ matrix.
In this section, we will present the expression for $\bf {L}$.
Following Rowe \cite{Rowe1}, we suppose that the studied system is described by the following general Hamiltonian with two-body interaction,
\begin{equation}
  H = \sum_{\nu \nu^{\prime}} T_{\nu \nu^{\prime}} a_{\nu}^{\dagger} a_{\nu^{\prime}} + \frac{1}{4} \sum_{\mu \nu}\sum_{\mu^{\prime} \nu^{\prime}} V_{\mu \nu \mu^{\prime} \nu^{\prime}} a_{\mu}^{\dagger} a_{\nu}^{\dagger} a_{\nu^{\prime}}a_{\mu^{\prime}}.    \label{Eq58} 
\end{equation}
For the moment, we assume that the above Hamiltonian are written on natural orbitals. The orbital update will be discussed in the next subsection. So we have $\langle a_{\nu}^{\dagger} a_{\nu^{\prime}} \rangle = \delta_{\nu \nu^{\prime}}\langle a_{\nu}^{\dagger} a_{\nu} \rangle$.

We can assign some constraints for the coefficients $T_{\nu \nu^{\prime}}$ and $V_{\mu \nu \mu^{\prime} \nu^{\prime}}$. The Hermiticity of $H$ can be guaranteed if we require 
\begin{eqnarray}
   && T_{\nu \nu^{\prime}} = T_{\nu^{\prime} \nu }^{\ast}, \nonumber \\
   && V_{\mu \nu \mu^{\prime} \nu^{\prime}} = V_{\mu^{\prime} \nu^{\prime}\mu \nu }^{\ast}.    \label{Eq59} 
\end{eqnarray}
We can require the following symmetry to $V_{\mu \nu \mu^{\prime} \nu^{\prime}}$ due to the anti-commutating properties of Fermions, 
\begin{eqnarray}
   && V_{\mu \nu \mu^{\prime} \nu^{\prime}} =- V_{\nu\mu  \mu^{\prime} \nu^{\prime}} , \nonumber \\
   && V_{\mu \nu \mu^{\prime} \nu^{\prime}} = - V_{\mu\nu  \nu^{\prime}\mu^{\prime} }.    \label{Eq60} 
\end{eqnarray}
Below, for simplicity, we will assume that $V_{\mu \nu \mu^{\prime} \nu^{\prime}}$ is real.

The Liouville matrix $\bf{L}$ in the form of Eq.(\ref{Eq26}) can be expressed in terms of the one- and two-particle density matrices.
Denoting $A_i = a_{\alpha}^{\dagger}a_{\beta}$ and  $A_j = a_{\alpha^{\prime}}^{\dagger}a_{\beta^{\prime}}$, we have
\begin{eqnarray}
&& (L_{11})_{ij}  \nonumber \\
&=& (A_i | [A_j, H])  = \langle [a_{\beta}^{\dagger}a_{\alpha}, [a_{\alpha^{\prime}}^{\dagger}a_{\beta^{\prime}}, H]] \rangle  \nonumber\\
&=& (\delta_{\beta \beta^{\prime}} T_{\alpha \alpha^{\prime}} - \delta_{\alpha \alpha^{\prime}} T_{\beta^{\prime} \beta})(n_{\alpha} - n_{\beta})  \nonumber \\
&& + \frac{1}{2} \delta_{\alpha \alpha^{\prime}}  \sum_{\nu \mu^{\prime} \nu^{\prime}} V_{\beta^{\prime} \nu \mu^{\prime} \nu^{\prime}} \langle a_{\beta}^{\dagger} a_{\nu}^{\dagger} a_{\nu^{\prime}} a_{\mu^{\prime}} \rangle \nonumber \\
&& + \frac{1}{2} \delta_{\beta \beta^{\prime}}  \sum_{\nu \mu^{\prime} \nu^{\prime}} V_{\alpha^{\prime} \nu \mu^{\prime} \nu^{\prime}} \langle a_{\alpha}^{\dagger} a_{\nu}^{\dagger} a_{\nu^{\prime}} a_{\mu^{\prime}} \rangle^{\ast} \nonumber \\
&& - \frac{1}{2}  \sum_{\mu \nu} \left[ V_{\mu \nu \beta^{\prime} \alpha} \langle a_{\beta}^{\dagger} a_{\alpha^{\prime}}^{\dagger} a_{\nu} a_{\mu} \rangle + V_{\mu \nu \alpha^{\prime} \beta} \langle a_{\alpha}^{\dagger} a_{\beta^{\prime}}^{\dagger} a_{\nu} a_{\mu} \rangle^{\ast} \right]   \nonumber \\
&& - \sum_{\nu \nu^{\prime}} \left[ V_{\beta \nu^{\prime} \beta^{\prime} \nu } \langle a_{\alpha^{\prime}}^{\dagger} a_{\nu}^{\dagger} a_{\nu^{\prime}} a_{\alpha} \rangle + V_{\alpha \nu^{\prime} \alpha^{\prime} \nu} \langle a_{\beta^{\prime}}^{\dagger} a_{\nu}^{\dagger} a_{\nu^{\prime}} a_{\beta} \rangle^{\ast} \right]   \nonumber \\
&&    \label{Eq61} 
\end{eqnarray}
Note that the Hermiticity $(L_{11})_{ij} = (L_{11})_{ji}^{\ast}$ should hold. This is a nontrivial requirement and it is equivalent to the identity $\langle [[A_i^{\dagger}, A_j], H] \rangle =0$. In the actual calculation, if additional approximations are introduced to calculate the averages, this equation may not hold.
Therefore, one needs to enforce the Hermiticity in the calculation by replacing $(L_{11})_{ij}$ with $\left[ (L_{11})_{ij} + (L_{11})_{ji}^{\ast} \right]/2$.
For $(L_{12})_{ij}$, we have
\begin{eqnarray}
   && (L_{12})_{ij} \nonumber \\
  && =  (A_i | [A_j^{\dagger}, H])  = \langle [a_{\beta}^{\dagger}a_{\alpha}, [a_{\beta^{\prime}}^{\dagger}a_{\alpha^{\prime}}, H]] \rangle  \nonumber\\
  && = (L_{11})_{ij} (\alpha^{\prime}  \rightarrow \beta^{\prime}, \, \beta^{\prime} \rightarrow \alpha^{\prime}).    \label{Eq62} 
\end{eqnarray}

In the renormalized RPA, one uses the Hatree-Fock decoupling to reduce the two-particle density matrix into the product of one-particle densities. The latter is then calculated self-consistently. In such a scheme, the $(L_{11})_{ij}$ is approximated as
\begin{eqnarray}
  && (L_{11})_{ij} \nonumber \\
  && \approx (n_{\alpha} - n_{\beta}) \left[\delta_{\beta \beta^{\prime}} T_{\alpha \alpha^{\prime}} - \delta_{\alpha \alpha^{\prime}} T_{\beta^{\prime} \beta} \right]  \nonumber \\
  && + (n_{\alpha} - n_{\beta}) \left[\delta_{\alpha \alpha^{\prime}}\left( \sum_{\nu} V_{\beta \nu \nu \beta^{\prime}}  n_{\nu} \right)     + \delta_{\beta \beta^{\prime}} \left( \sum_{\nu} V_{\alpha \nu  \alpha^{\prime} \nu}  n_{\nu} \right) \right]  \nonumber \\
  &&  + V_{\beta \alpha^{\prime} \beta^{\prime} \alpha}(n_{\alpha} - n_{\beta})  (n_{\alpha^{\prime}} - n_{\beta^{\prime}} ).    \label{Eq63} 
\end{eqnarray}
$(L_{12})_{ij}$ is also given by Eq.(\ref{Eq62}).

\end{subsection}

\begin{subsection}{Orbital Update}

Since the basis $\mathcal{B}_1 = \{ a_{\alpha}^{\dag} a_{\beta} \}$ ($\alpha \neq \beta$) is not invariant upon the orbital rotation (not containing $a_{\alpha}^{\dagger}a_{\alpha}$), the result of PTA on basis $\mathcal{B}_1$ will in general depend on the orbital selection.
Usually, the original Hamiltonian is written down in terms of the creation and annihilation operators defined on pre-selected orbitals which are not the natural orbitals. 

In Ref. \cite{Schuck1}, the authors minimized the ground state energy obtained by RPA with respect to the rotation of orbitals. The obtained orbitals are assumed to be natural orbitals. It is not clear whether the minimization of ground state energy and being natural orbital are compatible to each other.
In Ref. \cite{Catara1}, Catara {\it etc.} used the Kohn-Sham orbitals and assumed that they are natural orbitals.

Here, we propose to solve the natural orbital iteratively using PTA. Suppose we have a Hamiltonian $H[ \{a_{\alpha} \}]$, such as Eq.(\ref{Eq58}), which is written on certain original orbitals. After the RPA calculation, we obtain the one-particle density matrix ${\bf \rho}_{\nu \nu^{\prime}} = \langle a_{\nu}^{\dagger}a_{\nu^{\prime}} \rangle$. We can diagonalize it using a unitary matrix $\bf{W}$ as ${\bf W}^{-1} {\bf \rho} {\bf W} = \text{diag}(n_1, n_2, ..., n_L)$. 
The annihilation operator $\tilde{a}_{\alpha}$ for the new orbital, on which ${\bf \rho}$ is diagonal, is then expressed in terms of the original operator $a_{\alpha}$ as 
\begin{equation}   \label{Eq64}
   \tilde{a}_{\alpha} = \sum_{\nu=1}^{L} a_{\nu} W_{\nu \alpha}.   
\end{equation}
The inverse transformation reads
\begin{equation}   \label{Eq65}
   a_{\alpha} = \sum_{\nu=1}^{L} \tilde{a}_{\nu} W^{\ast}_{\alpha \nu}.   
\end{equation}
The Hamiltonian $H$ can now be written in terms of $\tilde{a}_{\alpha}$ as
\begin{equation}
  H = \sum_{\nu \nu^{\prime}} \tilde{T}_{\nu \nu^{\prime}} \tilde{a}_{\nu}^{\dagger} \tilde{a}_{\nu^{\prime}} + \frac{1}{4} \sum_{\mu \nu}\sum_{\mu^{\prime} \nu^{\prime}} \tilde{V}_{\mu \nu \mu^{\prime} \nu^{\prime}} \tilde{a}_{\mu}^{\dagger} \tilde{a}_{\nu}^{\dagger} \tilde{a}_{\nu^{\prime}} \tilde{a}_{\mu^{\prime}},   \label{Eq66}
\end{equation}
where the parameters $\tilde{T}_{\nu \nu^{\prime}}$ and $\tilde{V}_{\mu \nu \mu^{\prime} \nu^{\prime}}$ are 
\begin{equation}
  \tilde{T}_{\nu \nu^{\prime}} = \sum_{\alpha \beta} W^{T}_{\nu \alpha} T_{\alpha \beta} W^{\ast}_{\beta \nu^{\prime}},   \label{Eq67}
\end{equation}
and
\begin{equation}
  \tilde{V}_{\mu \nu \mu^{\prime} \nu^{\prime}} = \sum_{\alpha_1 \alpha_2} \sum_{\alpha_3 \alpha_4} W^{T}_{\mu \alpha_1}
  W^{T}_{\nu \alpha_2} V_{\alpha_1 \alpha_2 \alpha_3 \alpha_4} W^{\ast}_{\alpha_3 \mu^{\prime}}
  W^{\ast}_{\alpha_4 \nu^{\prime}} .   \label{Eq68}
\end{equation}
Note that $\tilde{T}_{\nu \nu^{\prime}}$ and $\tilde{V}_{\mu \nu \mu^{\prime} \nu^{\prime}}$ have the same symmetry as Eqs.(\ref{Eq59}) and (\ref{Eq60}). The updated Hamiltonian Eq.(\ref{Eq66}) has the same form as Eq.(\ref{Eq58}), but with updated orbitals and operators.

Starting from the original orbital, the orbitals are updated iteratively in the above procedure until the they converge to the true natural orbital (within RPA). In this process, the form of the physical quantity $\hat{O}(\{ a_{\alpha} \})$ must also be transformed and its average will converge as the orbitals convergence.

\end{subsection}

\begin{subsection}{Self-consistent Calculation of $\bf{I}$ and $\bf{L}$}

The matrices ${\bf I}$ and ${\bf L}$ contain the one-particle density $\langle a^{\dagger}_{\alpha} a_{\beta} \rangle$ and the two-particle density $\langle a^{\dagger}_{\alpha} a_{\beta}^{\dagger} a_{\gamma} a_{\delta} \rangle$. In a fully self-consistent RPA, they need to be calculated self-consistently from the GFs via the spectral theorem. The first problem of the self-consistent calculation is that the averages $\langle (A_{i})_d^{\dagger}(A_{j})_d \rangle$, $\langle (A_{i})_d^{\dagger}(B_{j})_d \rangle$, $\langle (B_{i})_d^{\dagger}(A_{j})_d \rangle$, and $\langle (B_{i})_d^{\dagger}(B_{j})_d \rangle$ produced by the spectral theorem do not contain the static contribution. The second problem is that they may not cover all the averages appearing in  $\bf{I}$ and $\bf{L}$. In particular, the one-particle density $\langle a_{\alpha}^{\dagger} a_{\beta} \rangle$ is not covered. For the first problem, we have used the approximation $\langle (A_{i})_0^{\dagger} (A_{j})_0 \rangle \approx 0$. In this part, we will address the second problem. 

What the spectral theorem directly produces is the matrix $\bf{C}$ of the form
\begin{equation}    \label{Eq69}
  {\bf C} = \left(
\begin{array}{cc}
C_{11} &  C_{12}  \\
C_{21} &  C_{22} \\
\end{array}
\right).
\end{equation}
Here, $(C_{11})_{ij} = \langle (A_{i})_d^{\dagger} (A_{j})_d \rangle$, $(C_{12})_{ij} = \langle (A_{i})_d^{\dagger}(B_{j})_d \rangle$, $(C_{21})_{ij} = \langle (B_{i})_d^{\dagger}(A_{j})_d \rangle$, and  $(C_{22})_{ij} = \langle (B_{i})_d^{\dagger}(B_{j})_d \rangle$.
The basis of the present formalism is sufficiently large such that $\bf{C}$ already contains most of the elements of the two-particle density matrix. For example, for the operators with four different subscripts ($\alpha, \beta, \gamma, \delta$), we have
\begin{equation}
 \langle a_{\alpha}^{\dagger}a_{\beta} a_{\gamma}^{\dagger}a_{\delta} \rangle 
= \langle (a_{\beta}^{\dagger}a_{\alpha})^{\dagger} ( a_{\gamma}^{\dagger} a_{\delta} ) \rangle.   \label{Eq70}
\end{equation}
For the operators with three different subscripts ($\alpha, \beta, \gamma$), we have
\begin{equation}
\langle a_{\alpha}^{\dagger}a_{\beta} a_{\gamma}^{\dagger}a_{\gamma} \rangle  
= \langle a_{\alpha}^{\dagger} a_{\beta} \rangle - \langle (a_{\gamma}^{\dagger}a_{\alpha})^{\dagger} (a_{\gamma}^{\dagger}a_{\beta} )\rangle.   \label{Eq71}
\end{equation}
For the operator with two different subscripts ($\alpha, \beta$), we have
\begin{equation}
   \langle a_{\alpha}^{\dagger}a_{\alpha} a_{\beta}^{\dagger}a_{\beta} \rangle 
= \langle a_{\alpha}^{\dagger}a_{\alpha} \rangle - \langle (a_{\beta}^{\dagger}a_{\alpha})^{\dagger} (a_{\beta}^{\dagger}a_{\alpha} )\rangle.   \label{Eq72}
\end{equation}
Both the above two equations involve one-particle density, which is not directly produced by the spectral theorem. One has to express $\langle a_{\alpha}^{\dag} a_{\beta} \rangle$ in terms of $\langle A_{i}^{\dag}A_{j} \rangle$, etc. 

At this point, we have different methods for calculating $\langle a_{\alpha}^{\dagger}a_{\beta} \rangle$. Note that the formalism obtained from PTA on the basis $\{a_{\alpha}^{\dagger}a_{\beta} \}$ ($\alpha \neq \beta$) applies both to the canonical ensemble and the grand canonical ensemble because the basis operator $a_{\alpha}^{\dagger}a_{\beta}$ commute with $\hat{N}$ (see Appendix B). But the additional method for calculating the one-particle density could apply either to the canonical ensemble or the grand canonical ensemble, making the whole self-consistent scheme applicable only to a specific ensemble. 
 
A frequently used method for calculating $\langle a_{\alpha}^{\dagger}a_{\beta} \rangle$ is the
number operator method of Rowe, which applies only to the canonical ensemble. Therefore, it is legitimate to combine PTA formula with the number operator method for canonical ensemble calculation.
In this paper, we propose the following exact formula based on number operator method, 
\begin{equation}    \label{Eq73}
   \langle a_{\alpha}^{\dagger}a_{\beta} \rangle = \frac{1}{L-N-1} \sum_{\gamma \neq \alpha, \beta} \langle (a_{\gamma}^{\dagger} a_{\alpha})^{\dagger} ( a_{\gamma}^{\dagger}a_{\beta} )\rangle \,\,\,\,(\alpha \neq \beta), 
\end{equation}
and
\begin{equation}     \label{Eq74}
   \langle a_{\alpha}^{\dagger}a_{\alpha} \rangle = \frac{1}{L-N} \sum_{\gamma \neq \alpha} \langle (a_{\gamma}^{\dagger} a_{\alpha})^{\dagger} ( a_{\gamma}^{\dagger}a_{\alpha} )\rangle .
\end{equation}
Here, $L$ is the number of orbitals. $N$ is the number of electrons of the system. The derivation of these equations are given in Appendix D.

In the grand canonical ensemble, the number operator method does not apply. In that case, $\langle a_{\alpha}^{\dagger}a_{\beta} \rangle$ can be calculated from an independent PTA calculations on the basis $\{ a_{\alpha}, a_{\alpha}^{\dagger}a_{\beta}a_{\gamma}, ... \}$, or $\{ a_{\alpha} a_{\beta} \}$. For the latter, PTA can produce the
averages of the type $\langle (a_{\alpha} a_{\beta})^{\dagger} (a_{\gamma}a_{\delta}) \rangle$, which are combined with Eq.(\ref{Eq71}) to produces
\begin{equation}
   \langle a_{\alpha}^{\dagger}a_{\beta} \rangle  = \langle (a_{\gamma}^{\dagger} a_{\alpha})^{\dagger} (a_{\gamma}^{\dagger} a_{\beta}) \rangle + \langle (a_{\gamma}a_{\alpha})^{\dagger} (a_{\gamma}a_{\beta}) \rangle.   \label{Eq75}
\end{equation}
Here, $\gamma$ is an arbitrary orbital other than $\alpha$ and $\beta$. In this paper, we mainly focus on the canonical ensemble and will not present the details of this scheme.

In this context, we would like to briefly comment on 
the difference between the sc-RPA scheme presented in this work and those developed within the realm of first-principles RPA calculations. As mentioned in the introduction,  RPA can be
interpreted as an orbital-dependent functional in the framework Kohn-Sham (KS) DFT. Within the strict KS scheme, the sc-RPA scheme amounts to determining a local, optimized effective potential (OEP) on which the RPA total energy is stationary 
. It has been demonstrated that such a local XC effective potential displays significantly improved behavior compared to
the conventional local and semilocal approximations \cite{Bleiziffer2013,Hellgren2015,Goerling2025}. Alternatively, one can introduce a non-local effective potential, within the generalized KS scheme, to optimize the RPA energy
\cite{Jin2017,Voora2019,Yu2021}. However, in this case, such a nonlocal effective potential is not uniquely defined. Either way, in the current sc-RPA schemes for first-principles calculations, only the single-particle orbitals and orbital energies are updated, but not the occupation numbers. In contrast, in the sc-RPA scheme for model Hamiltonian as described above, the occupation numbers are updated during the
self-consistency loop and become \textit{de facto} fractional, reflecting the correlated nature of the system. Introducing the fractional occupations in first-principles RPA calculations awaits for further investigations and the possible consequence is yet to be explored. 

\end{subsection}

\begin{subsection}{Correspondence with Rowe's Formalism}

In this part, we will transform the formalism developed above into the form of RPA in Rowe's symbols. We will show that in the limit $T=0$, Rowe \cite{Rowe1,Rowe2,Schuck1} and Catara's formalism \cite{Catara1} are recovered.

Rowe's formula is based on the normalized basis operators $\tilde{A}_{i}$ and $\tilde{B}_{i}$ as ($i = (\alpha > \beta)$ )
\begin{eqnarray}
  && \tilde{A}_{i} = \frac{A_i}{ \sqrt{(A_i|A_i)}} = \frac{a_{\alpha}^{\dagger} a_{\beta}}{\sqrt{n_{\beta} - n_{\alpha}}},  \nonumber \\
    && \tilde{B}_{i} = \frac{B_i}{ \sqrt{-(B_i|B_i)}} = \frac{a_{\beta}^{\dagger} a_{\alpha}}{\sqrt{n_{\beta} - n_{\alpha}}}.     \label{Eq76}    
\end{eqnarray}
Note that in the degenerate case with $n_{\alpha} = n_{\beta}$ for some $\alpha$ and $\beta$, Eq.(\ref{Eq76}) cannot be applied. In contrast, our PTA formalism still works under the help of zero-removing technique \cite{Jia1} or with a positive definite inner product.

We define $\bf{X}$ and $\bf{Y}$ matrices as
\begin{eqnarray}
  && {\bf X} \equiv \sqrt{I_{11}} U_{11},  \nonumber \\
  && {\bf Y} \equiv - \sqrt{I_{11}} U^{\ast}_{12}.       \label{Eq77}    
\end{eqnarray}
The first equation of Eq.(\ref{Eq31}) then becomes
\begin{eqnarray}
  O_{1\nu} &=& \sum_{i=1}^{D/2} \left[ X_{i \nu} (\tilde{A}_{i})_d - Y_{i \nu} (\tilde{B}_{i})_d \right] \nonumber \\
      & \approx &\sum_{i=1}^{D/2} \left[ X_{i \nu} \tilde{A}_{i} - Y_{i \nu} \tilde{B}_{i} \right] \nonumber \\
      &=& Q_{\nu}^{\dagger}.     \label{Eq78}    
\end{eqnarray}
Here, $Q_{\nu}^{\dagger}$ is the excitation operator introduced in Rowe's work \cite{Rowe1}. $\lambda_{1\nu} < 0$ corresponds to raising energy when $O_{1\nu}$ is applied to an eigenstate, being consistent with the assumption in Rowe's formula that $Q_{\nu}^{\dagger}|0 \rangle = | \nu \rangle$ at $T=0$. Here and below, the approximation Eq.(\ref{Eq47}) will be used, i.e., $(A_{i})_0 \approx 0$ and $(B_{i})_0 \approx 0$. Note that here the formula is for arbitrary temperature.

In Rowe's formalism, the average excitation energy is described by two matrices ${\bf A}$ and ${\bf B}$, whose elements are defined as $A_{ij} \equiv  \langle \left[ \tilde{ B} _i, [H, \tilde{A}_j ] \right] \rangle$ and $ B_{ij} \equiv  - \langle \left[ \tilde{B} _i, [H, \tilde{B}_j ] \right] \rangle$. They are obtained as
\begin{eqnarray}
  \bf{A} &=&  - I_{11}^{-1/2} L_{11} I_{11}^{-1/2},  \nonumber \\
  \bf{B} &=&  I_{11}^{-1/2} L_{12} I_{11}^{-1/2}.     \label{Eq79}    
\end{eqnarray}
Note that the negative semidefiniteness of $L_{11}$ is consistent with the positive semidefiniteness of ${\bf A}$ matrix. Inserting these equations into Eq.(\ref{Eq50}), we obtain the standard form of Rowe as
\begin{eqnarray}
&&   \left(
\begin{array}{cc}
{\bf A} &  {\bf B}  \\
{\bf B}^{\ast} &  {\bf A}^{\ast}  \\
\end{array}
\right)   \left(
\begin{array}{cc}
{\bf X} &  {\bf Y}^{\ast}  \\
{\bf Y} &  {\bf X}^{\ast}  \\
\end{array}
\right)   \nonumber \\
 &=&   \left(
\begin{array}{cc}
\bf{1} &  \bf{0}  \\
\bf{0} &  -\bf{1}  \\
\end{array}
\right) \left(
\begin{array}{cc}
{\bf X} &  {\bf Y}^{\ast}  \\
{\bf Y} &  {\bf X}^{\ast}  \\
\end{array}
\right)\left(
\begin{array}{cc}
{\bf \Omega} &  \bf{0}  \\
\bf{0} &  - {\bf \Omega}  \\
\end{array}
\right).     \label{Eq80}    
\end{eqnarray}
Here, ${\bf \Omega} = - \bf{
\Lambda}_{11} \geq 0$ is the excitation energy.

The generalized orthonormal relations Eq.(\ref{Eq51}) are expressed in terms of $\bf{X}$ and $\bf{Y}$ as
\begin{eqnarray}
   && \sum_{i=1}^{D/2} \left[ X_{i \nu}^{\ast} X_{i \nu^{\prime}} - Y_{i \nu}^{\ast} Y_{i \nu^{\prime}} \right] = \delta_{\nu \nu^{\prime}},  \nonumber \\
   && \sum_{i=1}^{D/2} \left[ X_{i \nu} Y_{i \nu^{\prime}} - Y_{i \nu} X_{i \nu^{\prime}} \right] = 0.     \label{Eq81}    
\end{eqnarray}
The closure relation Eq.(\ref{Eq52}) becomes
\begin{eqnarray}
&&   \sum_{\nu =1}^{D/2} \left[ X_{i\nu} X_{j \nu}^{\ast} - Y_{i\nu}^{\ast} Y_{j\nu}  \right] = \delta_{ij},    \nonumber \\
&&   \sum_{\nu =1}^{D/2} \left[ X_{i\nu} Y_{j \nu}^{\ast} - Y_{i\nu}^{\ast} X_{j\nu}  \right] = 0.    \label{Eq82}    
\end{eqnarray}

The expression of $A_{i}$ ($i = (\alpha > \beta)$) in terms of the eigen-excitation operators $O_{1\nu}$ and $O_{2\nu}$, Eq.(\ref{Eq54}), reads
\begin{equation}    \label{Eq83}
   A_{i} \approx (A_{i})_d = \sqrt{n_{\beta} - n_{\alpha}} \sum_{\nu=1}^{D/2} \left[ X_{i\nu}^{\ast} O_{1\nu} + Y_{i\nu} O_{2\nu} \right] .
\end{equation}
Combined with Eqs.(\ref{Eq32})-(\ref{Eq34}), Eq.(\ref{Eq83}) gives the averages like $\langle A_{i}^{\dagger} A_j \rangle$ as
\begin{eqnarray}
&&   \langle A_{i}^{\dagger} A_{j} \rangle   \nonumber \\
&\approx &  \sqrt{(I_{11})_{ii}(I_{11})_{jj}} \sum_{\nu=1}^{D/2} \left[ X_{i\nu}X_{j\nu}^{\ast} \frac{-1}{e^{\beta \lambda_{1\nu}} -1 } \right]  \nonumber \\
&& + \sqrt{(I_{11})_{ii}(I_{11})_{jj}} \sum_{\nu=1}^{D/2} \left[ Y^{\ast}_{i\nu} Y_{j\nu} \frac{1}{e^{-\beta \lambda_{1 \nu}} -1} \right].    \label{Eq84}
\end{eqnarray}

The spectral functions in Eqs.(\ref{Eq56}) and (\ref{Eq57}) reads in Rowe's symbol as
\begin{eqnarray}
&& \rho_{A_i, A_{j}^{\dagger}}(\omega) \nonumber \\
 &=& - \sqrt{(I_{11})_{ii}(I_{11})_{jj}} \sum_{\nu=1}^{D/2} \left[X^{\ast}_{i\nu}X_{j\nu}  \delta(\omega -\lambda_{1\nu}) \right] \nonumber \\
  &&  +\sqrt{(I_{11})_{ii}(I_{11})_{jj}} \sum_{\nu=1}^{D/2} \left[Y_{i\nu}Y^{\ast}_{j\nu}  \delta(\omega + \lambda_{1\nu}) \right], \nonumber \\
  &&    \label{Eq85}
\end{eqnarray}
and
\begin{eqnarray} 
&& \rho_{A_i, B_{j}^{\dagger}}(\omega) \nonumber \\
 &=& - \sqrt{(I_{11})_{ii}(I_{11})_{jj}} \sum_{\nu=1}^{D/2} \left[X^{\ast}_{i\nu} Y_{j\nu}  \delta(\omega -\lambda_{1\nu}) \right] \nonumber \\
  &&  +\sqrt{(I_{11})_{ii}(I_{11})_{jj}} \sum_{\nu=1}^{D/2} \left[Y_{i\nu} X_{j\nu}^{\ast}  \delta(\omega + \lambda_{1\nu}) \right]. \nonumber \\
  &&    \label{Eq86}
\end{eqnarray}

The above equations are identical to Rowe's formalism, except that they apply to arbitrary temperature $T$. As temperature goes to zero, considering that $\lambda_{1\nu} < 0$, Eqs.(\ref{Eq32})-(\ref{Eq34}) give
\begin{eqnarray}
   && \langle O_{1\nu}^{\dagger} O_{1\nu^{\prime}} \rangle_{T=0}  = \langle 0| Q_{\nu} Q_{\nu^{\prime}}^{\dagger} |0 \rangle\approx  \delta_{\nu \nu^{\prime}},      \label{Eq87} \\
   && \langle O_{2\nu}^{\dagger} O_{2\nu^{\prime}} \rangle_{T=0}  = \langle 0| Q_{\nu}^{\dagger} Q_{\nu^{\prime}} |0 \rangle \approx 0,      \label{Eq88} \\
      && \langle O_{1\nu}^{\dagger} O_{2\nu^{\prime}} \rangle_{T=0} = \langle 0| Q_{\nu} Q_{\nu^{\prime}} |0\rangle \approx  0.    \label{Eq89}
\end{eqnarray}
 Correspondingly, from Eq.(\ref{Eq84}), the average like $\langle A_{i}^{\dagger} A_j \rangle$ becomes
\begin{eqnarray}
&&   \langle A_{i}^{\dagger} A_{j} \rangle_{T=0} \approx  \sqrt{(I_{11})_{ii}(I_{11})_{jj}}\sum_{\nu=1}^{D/2} \left[ X_{i\nu}X_{j\nu}^{\ast} \right].     \label{Eq90}
\end{eqnarray}
These are consistent with Rowe's formula for the ground state average.
Therefore, the formalism that we present in this section can be regarded as a finite temperature extension of Rowe's formalism. Previous finite temperature RPA employs Matsubara GF \cite{Schuck2}. Our work shows that Rowe's formalism can be directly extended to finite $T$ for real frequency GF.

\end{subsection}

\end{section}

\begin{section}{Variants of RPA formalism }

\begin{subsection}{RPA on the restricted basis }

The above formalism is based on the basis $\{ a_{\alpha}^{\dagger}a_{\beta} \}$ ($\alpha \neq \beta$) which has a dimension
of $D=L(L-1)$. In the weak coupling case and low temperature, the ground state is close to a Fermi sea, Not all the basis operators are equally important since the relevant excitations are those that excite a particle from below the Fermi energy to above. Therefore, in many works, people use a basis where the operators excite a fermion from the occupied state below Fermi energy to the empty state above Fermi energy. For the sake of self-consistency, their Hermitian conjugate operators are also included into the basis. Here we give a brief account of the PTA-based formalism on restricted basis.

The restricted basis $\mathcal{B}_2$ is defined as
\begin{eqnarray}
\mathcal{B}_2 &=& \{ a_{m}^{\dagger}a_{i}, \,\, (a_{m}^{\dagger}a_{i})^{\dagger} \} \,\,\,\,\,\,(1 \leq i \leq N, \,\, N+1 \leq m \leq L). \nonumber \\
&&     \label{Eq91}
\end{eqnarray}
Here, the orbitals are labelled in descending order of occupation as before. 
$N$ is the number of electrons and $L$ is the total number of one-particle orbitals. $L$ is determined by the truncation scheme for a given system size. 
This basis has a dimension $D = 2N(L-N)$, smaller than $D=L(L-1)$ of basis $\mathcal{B}_{1}$. Note that the number of electrons $N$ is fixed for basis $\mathcal{B}_2$. Hence this basis is only suitable for the canonical ensemble.

As in the traditional RPA, what we have in mind is that the ground state is approximately a Slater determinant composed of $N$ occupied one-particle orbitals, denoted by $i$ or $j$ ($1 \leq i,j \leq N$). The unoccupied orbitals are denoted by $m$, $n$, $p$ and $q$ ($N+1 \leq m,n,p,q \leq L$). RPA produces the low lying particle-hole as well as the collective excitation energies and operators. By self-consistently computing $\bf{I}$ and $\bf{L}$, the ground state will be improved by incorporating particle-hole fluctuations around the Slater determinant state. 

The PTA formalism on basis $\mathcal{B}_2$ is the same as that for $\mathcal{B}_1$, except that due to the limited number of basis operators in $\mathcal{B}_2$, it is more difficult to produce all the averages appearing in $\bf{I}$ and $\bf{L}$ from the spectral theorem. So the self-consistent part will be different from that on $\mathcal{B}_1$ basis. It requires more additional approximations, usually built on the smallness of higher order hole-hole (particle-particle) correlations below (above) Fermi energy, or the smallness of the amplitude $Y$ \cite{Rowe1,Catara1}. 

The PTA equations on basis $\mathcal{B}_2$ can be obtained by the following substitution in the formalism for basis $\mathcal{B}_1$,
\begin{eqnarray}
   &&  \mathcal{B}_1 \,\, \text{basis}  \, \,\,\, \,\,\,\,\,\,\,\, \longrightarrow \,\,\,\,  \mathcal{B}_2 \,\, \text{basis}  \nonumber \\
   &&  i = (\alpha > \beta)  \,\,\,\, \longrightarrow \,\,\,\, k = (m,i)   \nonumber \\
   &&  j = (\alpha^{\prime} > \beta^{\prime}) \,\,\,\, \longrightarrow \,\,\,\, k^{\prime} =(n,j)   \label{Eq93}
\end{eqnarray}
Here, $1 \leq i,j \leq N$, $N+1 \leq m, n \leq L$ in $\mathcal{B}_2$ basis.
Smaller basis size usually means lower accuracy in the PTA results. Considering that RPA on the restricted basis $\mathcal{B}_2$ is less accurate than that on $\mathcal{B}_1$, and there have been abundant details in the literature (see the work of Catara \cite{Catara1}), in this paper, we will not dwell on the development of PTA-based formalism on this basis.

\end{subsection}

\begin{subsection}{Other variants of RPA}

Our PTA-based formalism provides a general framework to implement different RPA variants (for reviews, see \cite{Co1,Li3}).
Besides the fully self-consistent RPA discussed above, further simplifications can be introduced to calculate $\bf {I}$ and $\bf{L}$. In the renormalized RPA, one calculates the two-particle density using Hartree-Fock decoupling approximation, as done in Eq.(\ref{Eq63}), and self-consistently determine the remaining one-particle density matrix. 

The crudest approximation is to keep the Hartree part of  $\bf{L}$ and evaluate the remaining  one-particle density on a Hartree-Fock ground state. This leads to the so-called standard RPA. 

If the size of the basis is reduced to half of the full size and the basis is composed of excitation operators only, i.e., $B_{td} =\{ a_{\alpha}^{\dagger} a_{\beta}\}$  ($\alpha > \beta$), the PTA on this basis leads to the Tamm-Dancoff approximation.

The sc-RPA can also be extended to account for higher order correlations. This can be achieved by incorporating operators of the type $a_{\alpha}^{\dagger} a_{\beta}^{\dagger}a_{\gamma}a_{\delta}$ in the basis. This so-called second RPA \cite{Papakonstantinou1,Schuck4} naturally goes beyond the sc-RPA. But the practical calculation would be very complicated, due to the larger basis dimension and the complex expressions of $\bf{I}$ and $\bf{L}$.

\end{subsection}

\end{section}

\begin{section}{sc-RPA for one-dimensional spinless fermion model}

In this section, to demonstrate the implementation of the sc-RPA and benchmark its features, we apply it to the one-dimensional spinless fermion model. This model is one of the best studied many-body fermion systems with abundant nontrivial physical properties, such as the singularity of the occupation distribution near the Fermi momentum \cite{Ejima1}, asymptotic power law decay of the density-density correlation function \cite{Lukyanov1,Hikihara1,Ruhman1} and the evolving of the density spectral function with interaction \cite{Caux1,Caux2,Zhong1}. We compare the results of sc-RPA with those from exact diagonalization, Bethe Ansatz \cite{Schneider1}, and bosonization \cite{Lukyanov1}, to evaluate the strength and weakness of the present sc-RPA.

\begin{subsection}{Model Hamiltonian and sc-RPA equations}

The Hamiltonian of the spinless fermion model reads
\begin{equation}        \label{Eq6-1}
   H = -t \sum_{j=1}^{L} \left[ c_{j}^{\dagger} c_{j+1} e^{i\Delta} + c_{j+1}^{\dagger} c_{j} e^{-i\Delta}\right] 
   + V \sum_{j=1}^{L} n_{j}n_{j+1}.
\end{equation}
We use the periodic boundary condition $c_{L+1} = c_{1}$. $t=1.0$ is set as the energy unit. $\Delta \neq 0$ corresponds to a flux applied to the charge of fermions. In this work, a tiny $\Delta \neq 0$ is applied to lift the degeneracy of single particle energy levels. 

Due to Jordan-Wigner transformation \cite{Jordan1}, at $\Delta=0$, this model is equivalent to the anisotropic Heisenberg model, also known as XXZ model, whose properties are well studied. At half filling, the ground state is a Luttinger liquid in the regime $|V| < 2t$ (disordered paramagnetic state for XXZ model). At $V=2t$, the ground state transits into a charge density wave state (antiferromagnetic long-range ordered state for XXZ model) which is stable in $V >2t$, through a Kosterlitz-Thouless type continuous transition. At $V = -2t$, the ground state transits into the cluster solid state (ferromagnetic long-range ordered state for XXZ model) which is stable in $V <2t$ \cite{Giamarchi1,Barghathi1}, through a first order phase transition.

Here, we demonstrate the sc-RPA algorithm by focusing on the parameter regime $|V| < 2t$ at half filling $N=L/2$, where the ground state is a Luttinger liquid \cite{Sachdev1}. The standard and renormalized RPA take the Fermi liquid state as the starting point, which has a well defined Fermi surface. They cannot produce the correct Luttinger liquid state if the ground state is not modified sufficiently through self-consistent calculation. Using this model, we wish to demonstrate to what extent the nontrivial ground state correlation can be captured by the present sc-RPA.

After the Fourier transformation 
\begin{equation}
c_{k}  = (1/\sqrt{L}) \sum_{j=1}^{L} c_j e^{-ijk},
\end{equation}
with $k= (2\pi/L) m_k$ ($m_k = 0, 1, ..., L-1$), the above Hamiltonian becomes
\begin{equation}       \label{Eq6-2}
   H = \sum_{k} T_{k}c_{k}^{\dagger} c_{k} + \frac{1}{4} \sum_{k_1, k_2, k_3, k_4} V_{k_1 k_2 k_3 k_4}  c_{k_1}^{\dagger} c_{k_2}^{\dagger} c_{k_4}c_{k_3}.
\end{equation}
The parameters here are
\begin{equation}       \label{Eq6-2.5}
    T_{k} = -2t \cos(k+\Delta)
\end{equation}
and 
\begin{eqnarray}       \label{Eq6-3}
&&    V_{k_1 k_2 k_3 k_4}   \nonumber \\
                 && = \frac{2V}{N} \left[ \cos(k_1 -k_3) - \cos(k_1 - k_4) \right] \delta_{k_1+k_2, k_3+k_4+2\pi n}. \nonumber \\
&&
 \end{eqnarray}
The interaction $V_{k_1 k_2 k_3 k_4}$ fulfils the symmetry relations Eqs.(\ref{Eq59}) and (\ref{Eq60}).
 For $\Delta =0$, a two fold degeneracy of single particle
energy $T_{k} = T_{2\pi -k}$ holds. This degeneracy will pose difficulty to the sc-RPA calculation since the inner product matrix $I_{k k^{\prime}}$ will have zero eigenvalues. A tiny $\Delta \neq 0$ can lift this degeneracy and stabilize the iterative sc-RPA calculation.

We use the particle-hole excitation operators as our basis. Due to the conservation of total momentum, the full basis is divided into $L$-dimensional subspaces of different momentum $q$, each containing operators
\begin{equation}       \label{Eq6-4}
   A^{q}_k = c_k^{\dagger} c_{k+q}  \,\,\,\,\,\,\,\, (m_k = 0, 1, 2, ..., L-1).
\end{equation}
We will omit the $q=0$ subspace since it is not required for establishing a closed set of equations for the averages. For this system, the momentum orbitals are natural orbitals. For $q \neq 0$, $(A^{q}_{k})_0 \neq 0$ if two degenerate eigenstates differ in the total momentum by $q$. Here, we use the approximation $(A^{q}_{k})_0 \approx 0$ ($q \neq 0$). Most significantly, $A^{q}_{k}$ and $(A^{q}_{k})^{\dagger} = A^{2\pi-q}_{k+q}$ belong to different subspaces and the matrices ${\bf L}$, ${\bf I}$ and ${\bf M}$ become block diagonal. We therefore write down the sc-RPA equations for each single block, instead of in the $2 \times 2$ block matrix form. 
For the inner product, we still use the commutator inner product $(X|Y) \equiv \langle [X^{\dag}, Y]\rangle$, which is conventional for sc-RPA.

The inner product matrix ${\bf{I}}^{q}$ is obtained as
\begin{equation}       \label{Eq6-5}
 I^{q}_{k k^{\prime}} = \delta_{k, k^{\prime}} \left( \langle n_{k+q} \rangle - \langle n_k \rangle \right).
\end{equation}
The matrix element of the Liouville matrix $ L^{q}_{k k^{\prime}} \equiv (A^{q}_{k} | [A^{q}_{k^{\prime}}, H] )$ is obtained as 
\begin{eqnarray}       \label{Eq6-6}
  L^{q}_{k k^{\prime}} &=& \delta_{k,k^{\prime}} \left(T_{k} - T_{k+q} \right) \left( \langle n_k \rangle - \langle n_{k+q}\rangle \right)  \nonumber \\
   && + \frac{1}{2} \delta_{k,k^{\prime}} \sum_{k_1 k_2 k_3} V_{k+q k_1 k_2 k_3} \langle c_{k+q}^{\dag} c_{k_1}^{\dag} c_{k_3} c_{k_2} \rangle   \nonumber  \\
   && + \frac{1}{2} \delta_{k,k^{\prime}} \sum_{k_1 k_2 k_3} V_{k k_1 k_2 k_3} \langle c_{k_2}^{\dag} c_{k_3}^{\dag} c_{k_1} c_{k} \rangle   \nonumber  \\
   && - \frac{1}{2} \sum_{k_1 k_2} V_{k_1 k_2 k^{\prime}+q k} \langle c_{k+q}^{\dag} c_{k^{\prime}}^{\dag} c_{k_2} c_{k_1} \rangle   \nonumber  \\
   && - \frac{1}{2} \sum_{k_1 k_2} V_{k_1 k_2 k^{\prime} k+q} \langle c_{k_1}^{\dag} c_{k_2}^{\dag} c_{k^{\prime}+q} c_{k} \rangle   \nonumber  \\
   && -  \sum_{k_1 k_2} V_{k+q k_2 k^{\prime}+q k_1} \langle c_{k^{\prime}}^{\dag} c_{k_1}^{\dag} c_{k_2} c_{k} \rangle   \nonumber  \\
   && - \sum_{k_1 k_2} V_{k k_2 k^{\prime} k_1} \langle c_{k+q}^{\dag} c_{k_2}^{\dag} c_{k_1} c_{k^{\prime}+q} \rangle.
\end{eqnarray}
Due to $(A^{q}_{k})_0 \approx 0$ for $q \neq 0$, Eq.(\ref{Eq20.2}) gives
\begin{equation}       \label{Eq6-7}
   {\bf C }^{q} \approx - {\bf I}^{q} \left( e^{\beta {\bf M}^{q} } - \bf{1} \right)^{-1}
\end{equation}
with ${\bf M}^{q} = [ {\bf{I}}^{q} ]^{-1} {\bf{L}}^{q}$ for each $q \neq 0$.

In order to complete the self-consistent loop, we need to express the averages involved in ${\bf{I}}^{q}$ and ${\bf{L}}^{q}$ in terms of ${\bf {C}}^{q}$.
For the averages of the type $\langle c_{\alpha}^{\dag} c_{\beta}^{\dag} c_{\gamma} c_{\delta} \rangle$ appearing in ${\bf{L}}^{q}$, we have (for subscripts fulfilling the momentum conservation $\alpha + \beta = \gamma + \delta + 2\pi m$, $m$ being an integer)
\begin{eqnarray}       \label{Eq6-8}
   \langle c_{\alpha}^{\dag}c_{\beta}^{\dag} c_{\gamma} c_{\delta} \rangle  \nonumber 
  &=& ( 1-\delta_{\alpha \beta} ) ( 1-\delta_{\gamma \delta} )  \left( \delta_{\beta\gamma } 
   \delta_{\alpha \delta}  - \delta_{\alpha \gamma} \delta_{\beta \delta } \right) \langle n_{\alpha} \rangle  \nonumber \\
  && +\left[  \delta_{\alpha \gamma} \left( 1 - \delta_{\alpha \delta} \right) C^{\alpha -\delta}_{\delta \beta}  - 
     \left( 1 - \delta_{\alpha \gamma} \right) C^{\alpha -\gamma}_{\gamma \beta} \right]   \times       \nonumber \\
  &&  (1- \delta_{\alpha \beta} ) (1- \delta_{\gamma \delta} )\nonumber \\.
\end{eqnarray}
For the fermion occupation number $\langle n_{k} \rangle$, it can be obtained from ${\bf C}$ either according to Eq.(\ref{Eq74}),
\begin{equation}       \label{Eq6-9}
   \langle n_{k} \rangle = \frac{1}{L-N} \sum_{k^{\prime} (\neq k)} C^{k - k^{\prime}}_{k^{\prime}k^{\prime}},
\end{equation}
or from 
\begin{equation}       \label{Eq6-10}
   \langle n_{k} \rangle = 1 - \frac{1}{N} \sum_{q (\neq 0)} C^{q}_{k k}.
\end{equation}
Both the above equations apply only to the canonical ensemble. They are the reducibility relations between the
one-body reduced density matrix (1RDM) and the two-body reduced density matrix (2RDM). 

The sc-RPA amounts to self-consistently solving the Eqs.(\ref{Eq6-5})-(\ref{Eq6-10}) for $\bf{C}$ and $\{ \langle n_k \rangle \}$. However, the solution can only be determined up to a global constant factor, which must be determined by the total fermion number constraints

\begin{equation}       \label{Eq6-11}
   \sum_k \langle n_k \rangle = N.
\end{equation}

\end{subsection}

\begin{subsection}{ Constraints from N-representability and Hamiltonian symmetry }

We carry out iterative solution of the above equations. It turns out that the naive iteration is stable only for $V$ very close to zero ($|V| \lesssim 0.05$). We find that enforcing the fermion exchange symmetry of the 2RDM \cite{Coleman1,DePrince1} can significantly enlarge the parameter regime where the iteration converges smoothly. Though this is a standard practice in the variational 2RDM methods \cite{Mazziotti1,Li1,Li2}, to our knowledge, it has not been incorporated into the sc-RPA calculations. In the practical calculations of this work, we have enforced the following relations of $\bf {C}$ elements.
By exchanging the positions of two annihilation operators in $C^{q}_{k k^{\prime}}$, we obtain 
\begin{equation}       \label{Eq6-12}
  C^{q}_{k k^{\prime}} = \delta_{q,0} \langle n_k \rangle + \delta_{k k^{\prime}} \langle n_{k+q} \rangle - C^{k-k^{\prime}}_{k^{\prime}+q \, k^{\prime}}.
\end{equation}
Similarly, exchanging two creation operators, we obtain
\begin{equation}       \label{Eq6-13}
  C^{q}_{k k^{\prime}} = \delta_{q,0} \langle n_{k^{\prime}} \rangle + \delta_{k k^{\prime}} \langle n_{k+q} \rangle - C^{k^{\prime}-k}_{k \, k+q}.
\end{equation}
By exchanging a pair of $c^{\dag}c$, we obtain
\begin{equation}       \label{Eq6-14}
  C^{q}_{k k^{\prime}} = \delta_{k k^{\prime}} \left( \langle n_{k+q} \rangle  - \langle n_k \rangle \right) 
                         +  C^{2\pi - q}_{k^{\prime}+q \, k+q}.
\end{equation}
In the above equations, the results of plus and minus of two momenta are understood as module $2\pi$. 
Eqs.(\ref{Eq6-12})-(\ref{Eq6-14}) belongs to the general N-representability relations fulfilled by 2RDM.
They are therefore the necessary conditions for ${\bf C}$ to be the 2RDM of a physical many-body thermal state.
One can prove that once the diagonal part of Eq.(\ref{Eq6-14}) and the total fermion number constraints Eq.(\ref{Eq6-11}) 
are fulfilled, the two reducibility relations Eq.(\ref{Eq6-9}) and (\ref{Eq6-10}) are equivalent.

Note that Eq.(\ref{Eq6-14}) can also be derived from the fact that the basis operators in the $q$ subspace is the Hermitian conjugate of the subspace $2\pi - q$, i.e., $\left( A^{q}_{k} \right)^{\dag}  = A^{2\pi-q}_{k+q} $. This relation gives constrains, besides Eq.(\ref{Eq6-14}), to other quantities of subspaces $q$ and $2\pi-q$. In particular, we have
\begin{eqnarray}       \label{Eq6-15}
&&  (I^{q})^{\ast}_{k k^{\prime}} = - I^{2\pi -q}_{k+q \, k^{\prime}+q},  \nonumber \\
&&  (L^{q})^{\ast}_{k k^{\prime}} = L^{2\pi -q}_{k+q \, k^{\prime}+q},  \nonumber \\  
&&  (M^{q})^{\ast}_{k k^{\prime}} = - M^{2\pi -q}_{k+q \, k^{\prime}+q}.
\end{eqnarray}
In our sc-RPA calculations, we have made use of these relations to reduce the number of independent variables.
Considering that all matrices above are real in our sc-RPA calculation, the last equation implies that ${\bf M}^{q}$
and ${\bf M}^{2\pi- q}$ has opposite eigen values, i.e., ${\bf \Lambda}^{q} = - {\bf \Lambda}^{2\pi-q}$

There are other constraints as well. For example, ${\bf C}$ is positive semi-definite (${\bf C} \succeq 0$), and ${\bf L}$ is negative semi-definite (${\bf L} \preceq 0$). These inequality constraints are more complicated to implement in the calculation. Therefore, in our sc-RPA calculation, we only monitor them but not enforce them. It turns out that in most situations where the iterative solution converges stably, ${\bf L} \preceq 0$ holds automatically, while ${\bf C} \succeq 0$ is sometimes slightly violated by a single smallest eigenvalue of the order $-10^{-3}$ . It seems that this slight violation of ${\bf C} \succeq 0$ does not have any significant influence on the quality of other observables such as $\langle n_k \rangle$ or correlation functions. The instability of iteration is always accompanied with the violation of ${\bf L} \preceq 0$.

Besides the constraints from N-representability discussed above, we have additional Hamiltonian symmetry besides the spatial translation symmetry.

\begin{subsubsection}{spatial inversion symmetry at $\Delta =0$}

For $\Delta =0$, the Hamiltonian is invariant with respect to the spatial inversion transformation, $c_k \rightarrow \tilde{c}_k = c_{2\pi -k}$. This transformation maps the basis operator $A^{q}_{k} \rightarrow \tilde{A}^{q}_k = A^{2\pi-q}_{2\pi -k}$. Due to the invariance of $H(\Delta=0)$ under this transformation, we have the following constraints at $\Delta=0$,
\begin{equation}       \label{Eq6-16}
 X^{q}_{k k^{\prime}} = X^{2\pi - q}_{2\pi -k \,\, 2\pi-k^{\prime}},   \,\,\,\,\,\,\,\, ({\bf X}={\bf I},{\bf M },{\bf L}, {\bf C} ). 
\end{equation}
This symmetry also leads to the relation $\langle n_k \rangle = \langle n_{2\pi -k} \rangle$ at $\Delta=0$.
\end{subsubsection}

\begin{subsubsection}{particle-hole symmetry at $\Delta =0$ and $N/L = 1/2$}

At $\Delta =0$ and half fermion filling $N/L = 1/2$, Hamiltonian Eq.(\ref{Eq6-1}) is invariant with respect to the particle-hole transformation, $c_i \rightarrow \tilde{c}_i = (-1)^{i+1} c_{i}^{\dag}$. In momentum space, it is written as $c_k \rightarrow \tilde{c}_{k} = - c_{\pi -k}^{\dag}$. This transformation maps the basis operator $A^{q}_{k} \rightarrow \tilde{A}^{q}_k = -A^{q}_{\pi -k-q}$. Due to the invariance of $H(\Delta=0)$ under this transformation at half filling, we have the following constraints,
\begin{equation}       \label{Eq6-17}
 X^{q}_{k k^{\prime}} = X^{q}_{\pi -k -q \,\, \pi-k^{\prime}-q},   \,\,\,\,\,\,\,\, ({\bf X}={\bf I},{\bf M },{\bf L}, {\bf C} ). 
\end{equation}
We also have the relation $\langle n_k \rangle + \langle n_{\pi-k} \rangle = 1$.

\end{subsubsection}

\begin{subsubsection}{combined inversion and particle-hole symmetry at $N/L = 1/2$ and arbitrary $\Delta$}

For arbitrary $\Delta$ and at half filling, although $H(\Delta)$ is not invariant under the above two transformations separately, it is invariant under the combined transformation $c_k \rightarrow \tilde{c}_{k} = - c_{k+ \pi}^{\dag}$. This transformation maps the basis operator $A^{q}_{k} \rightarrow \tilde{A}^{q}_k = - A^{-q}_{k+q+\pi}$. Due to this symmetry, we have the following constraints,
\begin{equation}       \label{Eq6-18}
 X^{q}_{k k^{\prime}} = X^{-q}_{k +q+\pi \,\, k^{\prime}+q+\pi},   \,\,\,\,\,\,\,\, ({\bf X}={\bf I},{\bf M },{\bf L}, {\bf C} ). 
\end{equation}
This symmetry leads to the relation $\langle n_k \rangle + \langle n_{k+\pi} \rangle = 1$.

In our sc-RPA calculation, we checked the above symmetry properties. We find that Eq.(\ref{Eq6-18}) is always obeyed. For Eqs.(\ref{Eq6-16}) and (\ref{Eq6-17}), we can only check them by taking the limit $\Delta \to 0^{+/-}$. It is found that for $N$ odd, the results are continuous at $\Delta=0$ and fulfil Eqs.(\ref{Eq6-16}) and (\ref{Eq6-17}). For $N$ even, due to the open shell effect, the $k$-resolved quantities such as $\langle n_k \rangle$ differs at $\Delta \to 0^{+}$ and $\Delta \to 0^{-}$, and do not fulfil Eqs.(\ref{Eq6-16}) and (\ref{Eq6-17}). But the $k$-integrated quantities such as  $\rho(q, \omega)$ are continuous at $\Delta=0$ and still fulfil them.

\end{subsubsection}

\end{subsection}

\begin{subsection}{sc-RPA algorithm}

In the sc-RPA calculation, we first choose an initial ${\bf C}$. Here, we use the $V=0$ results for the initialization, which
fulfils all the symmetry requirements automatically.
For a given ${\bf C}$, we first check the N-representability relations Eqs.(\ref{Eq6-12})-(\ref{Eq6-14}). If they are not fulfilled, symmetrize ${\bf C}$ accordingly. Then we produce $\{ \langle n_k \rangle \}$ from Eq.(\ref{Eq6-9}). Inserting ${\bf C}$ and $\{ \langle n_k \rangle \}$ in the the expressions Eqs.(\ref{Eq6-5}) and (\ref{Eq6-6}), we get ${\bf I}^{q}$ and ${\bf L}^{q}$ for each $q \neq 0$. From them, we solve the generalized eigen problem 
\begin{equation}       \label{Eq6-18.5}
 {\bf L}^{q} {\bf U}^{q} = {\bf I}^{q} {\bf U}^{q} {\bf \Lambda}^{q}
\end{equation}
and obtain ${\bf U}^{q}$ and ${\bf \Lambda}^{q}$. In this process, one can employ the $q \sim 2\pi-q$ symmetry Eq.(\ref{Eq6-15}) to save half of the computation cost. With the obtained ${\bf U}^{q}$ and ${\bf \Lambda}^{q}$, we produce the matrix ${\bf Q}^{q} = (e^{\beta {\bf M}^{q}}-{\bf 1})^{-1}$ that appears in Eq.(\ref{Eq6-7}). Fixing this ${\bf Q}^{q}$, we then solve the linear equations about ${\bf C}$ and $\{ \langle n_k \rangle \}$, i.e., Eqs.(\ref{Eq6-7}), (\ref{Eq6-9}) (or (\ref{Eq6-10})), and Eq.(\ref{Eq6-11}). The obtained ${\bf C}$ is then fed back to the beginning of the calculation and start a new iteration. This process is carried on until the difference between the input and output ${\bf C}$s are smaller than a given criterion. Note that ${\bf C}$ can also be directly calculated from the inverse transformation of Eq.(\ref{EqE8}), but it turns out that it leads to unstable iteration.
The two key steps in this process are solving the generalized eigenvalue problem and solving the linear equations for ${\bf C}$, respectively. Detailed algorithms are summarized in Appendix E.

\end{subsection}

\begin{subsection}{Results of sc-RPA}

In this part, we report the sc-RPA results for the one-dimensional spinless fermion model. We compare these results with those from ED, Bethe ansatz, and bosonization. The purpose is to show the advantage and weakness of sc-RPA in a model whose ground state is far away from the Fermi liquid and set a starting point for the future improvement.

\begin{figure}
 \vspace*{-1.5cm}
\begin{center}
  \includegraphics[width=280pt, height=230pt, angle=0]{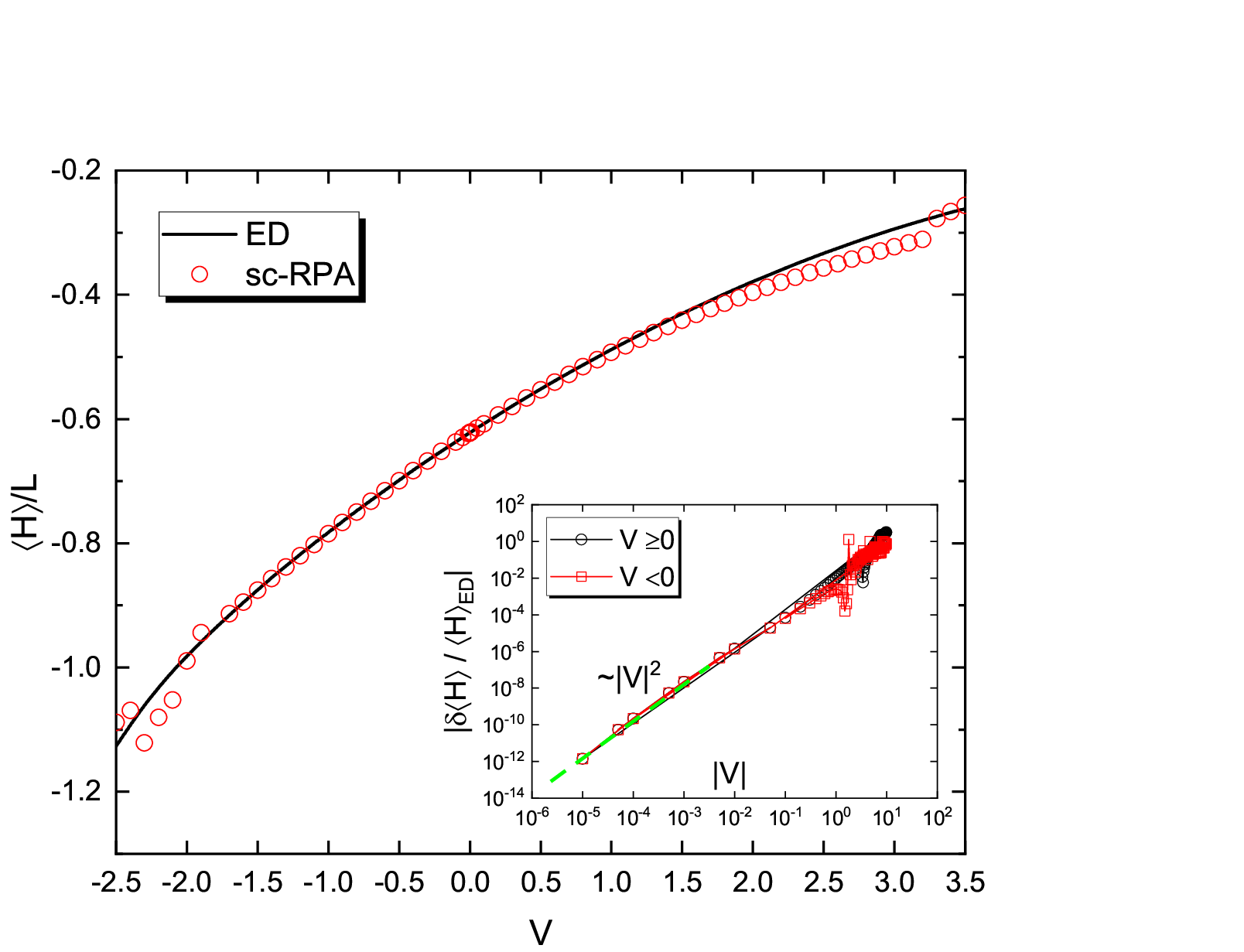}
  \vspace*{-1.0cm}
\end{center}
  \caption{(color online) Average energy per site as a function of $V$, obtained from sc-RPA (circles) and ED (solid line). Inset: the relative error of sc-RPA result with respect to ED, as functions of $|V|$ in the positive $V$ (circles) and negative $V$ (squares) regimes. The green dashed line is the eye-guiding line for the $V^{2}$ behaviour. Parameters are $L=12$, $N=6$, $t=1.0$, $T=10^{-4}$, $\Delta = 10^{-5}$, and the convergence precision of sc-RPA $10^{-7}$.
}   \label{Fig1}
\end{figure}

Figure 1 shows the average energy per site $\langle H \rangle/L$ as a function of $V$ for an $L=12$-site chain and $N=6$ at the low temperature $T=10^{-4}$. For this temperature, the properties are practically the same as that of the ground state. It is compared with the exact diagonalization (ED) result. It is seen that the sc-RPA energy is pretty good in the regime $-1.5 \lesssim V \lesssim 2.0$. The iteration becomes unstable in $V \lesssim -1.5$ and $V \gtrsim 3.0$. We observed that the numerical instability is associated with the breaking down of the negative definiteness of ${\bf L}$. 

The inset of Fig.1 shows the relative error of ground state energy. It increases as $V^{2}$ both on the positive and negative $V$ sides, showing that sc-RPA is accurate at $V^{1}$ level. At this level, the GF already contains the exchange and vertex correction diagrams. Note that the above comparisons are made at half filling. We have observed numerically that our sc-RPA gives exact results for the fermion number $N=0$, $N=1$, $N=L-1$, and $N=L$ at any $V$. 
This is expected because at these special fillings there is no correlation. 

\begin{figure}
 \vspace*{-2.0cm}
\begin{center}
  \includegraphics[width=310pt, height=230pt, angle=0]{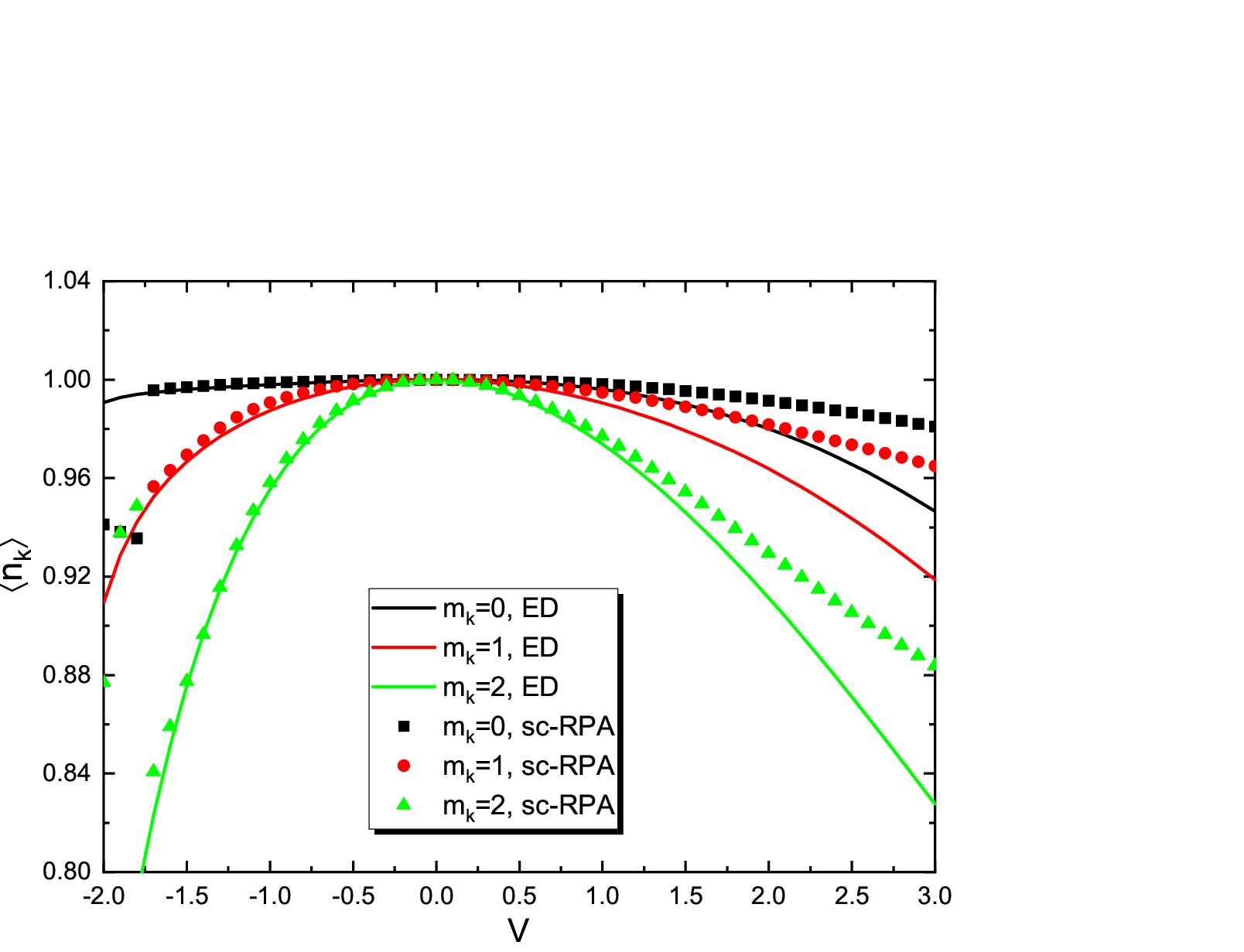}
  \vspace*{-1.0cm}
\end{center}
  \caption{(color online) Fermion occupation $\langle n_k \rangle$ of the momenta $k= 2\pi m_k /L + \Delta$ for $m_k = 0$, $1$, and $2$ (from top to bottom). They are obtained from sc-RPA (symbols) and ED (solid lines), respectively. Parameters are $L=12$, $N=6$, $t=1.0$, $T=0.01$, $\Delta = 0.2$, and the convergence precision of sc-RPA $10^{-7}$.
}   \label{Fig2}
\end{figure}

The correlation in the ground state is also reflected in the momentum space occupation number $\langle n_k \rangle$. In the Hartree-Fock ground state, $\langle n_k \rangle_{HF} = 0$ or $1$, depending on $T_{k}$ above or below $E_F$. The fractional occupation is thus a signature of correlation effect. In Fig.2, we show the distribution of $\langle n_k \rangle$ as functions of $V$ for the same chain as in Fig.1, and compare them with ED results. At $\Delta=0$ and half-filling, we have the symmetry $\langle n_{k} \rangle = 1 - \langle n_{\pi-k} \rangle$. In Fig.2, we show $\langle n_{k} \rangle(V)$ for $\Delta=0.2$, at which $\langle n_{k} \rangle \approx 1 - \langle n_{\pi-k} \rangle$. We therefore only show the results for the first three momenta. It is seen that for about $-1.5 \lesssim V \lesssim 1.0$, these occupations are quite close to ED results,  with fractional occupations up to $0.8$. 

In the regime $0 < |V| < 2t$, one expects the Luttinger liquid behavior $|\langle n_k \rangle  - \langle n_{k_F} \rangle| \sim |k-k_F|^{\frac{(1-2\eta)^{2}}{4\eta}}$ \cite{Ejima1} in the thermodynamic limit and $k \to k_F$. $\eta$ is given in Eq.(\ref{Eq7-2}). Due to the $L^4$ computational cost of the present sc-RPA, our practical calculation is limited to the chain of a few hundreds sites. We have not been able to approach the very vicinity of $k_F$ to observe such singularity. However, we do observe some features of a Luttinger liquid ground state in the correlation function of sc-RPA. See below.

\begin{figure}
 \vspace*{-2.0cm}
\begin{center}
  \includegraphics[width=420pt, height=330pt, angle=0]{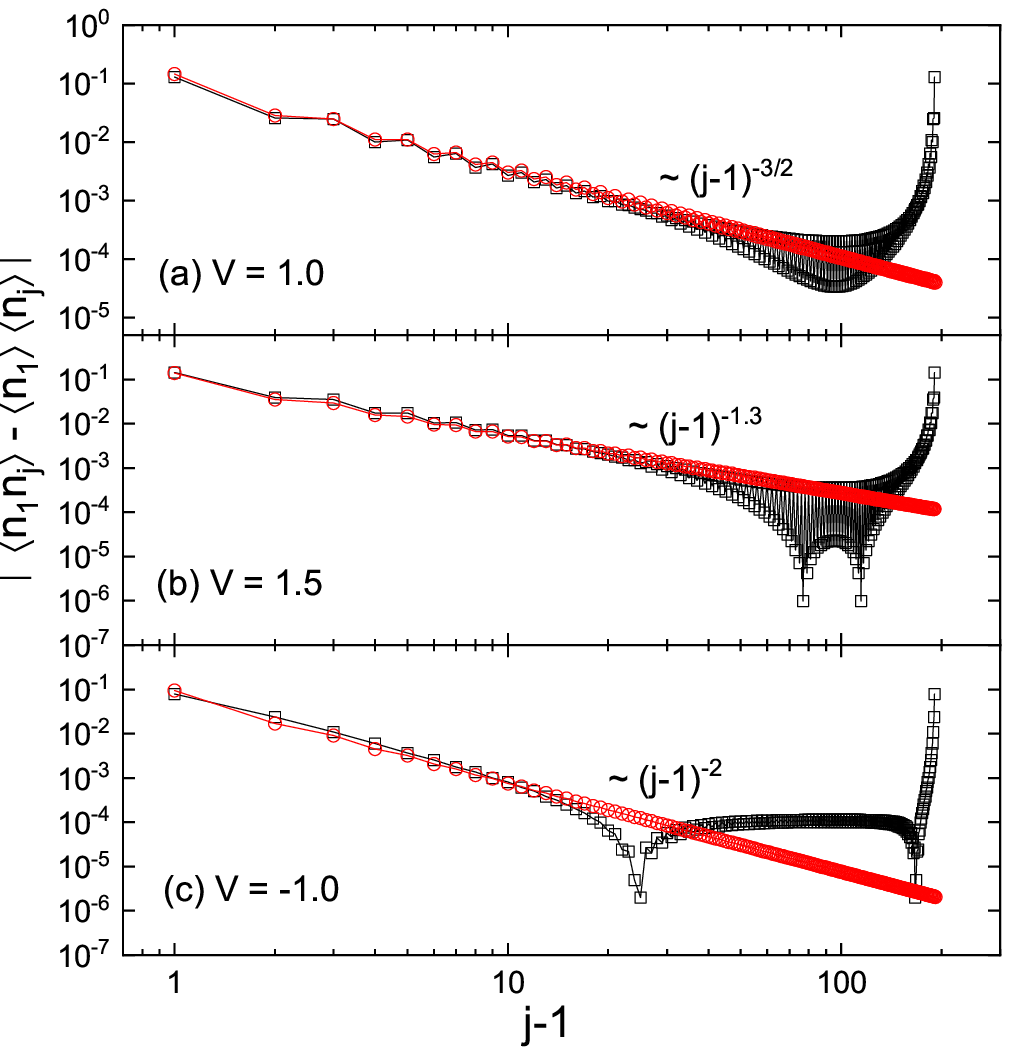}
  \vspace*{-1.0cm}
\end{center}
  \caption{(color online) Absolute value of density-density correlation function $|C_{1j}| = |\langle n_1 n_j \rangle - \langle n_1 \rangle \langle n_j \rangle|$ as functions of $j-1$. (a) $V=1.0$, (b) $V=1.5$, and (c) $V=-1.0$. They are obtained from sc-RPA (black squares) and bosonization \cite{Lukyanov1} (red circles). The lines are for guiding eyes. The results of sc-RPA are obtained at parameters $L=192$, $N=96$, $t=1.0$, $T=10^{-3}$, $\Delta=10^{-5}$, while those of bosonization are at $L=2N=\infty$, $t=1.0$, $T=0$, and $\Delta=0$. The long-distance asymptotic power from Eqs.(\ref{Eq7-1}) and (\ref{Eq7-2}) are marked in the figure.
   }   \label{Fig3}
\end{figure}

Fig.3 shows the absolute value of the density-density correlation function $C_{1j} \equiv \langle n_1 n_j \rangle - \langle n_1 \rangle \langle n_j \rangle$. obtained from sc-RPA (squares) for a finite chain of $L=192$ at half filling and a low temperature $T=10^{-3}$, with tiny momentum shift $\Delta=10^{-5}$. It is compared with the results from bosonization in the thermodynamic limit and at zero temperature. It is found that they agree remarkably well in the short-to-intermediate range where the finite-size effect is small.

In particular, the asymptotic power law decay from bosonization \cite{Lukyanov1} (exponent marked in the figure) reads
\begin{eqnarray}\label{Eq7-1}
C_{1j}(j \to \infty) &=& \frac{-1}{4\pi^{2} \eta} (j-1)^{-2} + \frac{A}{4}(-1)^{j-1} (j-1)^{-\frac{1}{\eta}} +...  \nonumber \\
&&
\end{eqnarray}
Here,
\begin{equation}  \label{Eq7-2}
\eta = \frac{1}{\pi} \arccos(-\frac{V}{2t}), 
\end{equation}
and
\begin{eqnarray}     \label{Eq7-2p}
A&=&\frac{8}{\pi^{2}}\left[\frac{\Gamma[\eta/(2-2\eta)]}{2\sqrt{\pi}[1/(2-2\eta)]}\right]^{1/\eta} \nonumber \\
&&\times \exp \left[\int_{0}^{\infty}\frac{dt}{t}\left(\frac{\sinh[(2\eta-1)t]}{\sinh(\eta t)\cosh[(1-\eta)t]}\right. \right.\nonumber \\
&&\left.\left. -\frac{2\eta-1}{\eta}e^{-2t}\right)\right].
\end{eqnarray}

From this expression, $C_{1j} \sim (j-1)^{-2}$ for $-2t < V \leq 0$ and $C_{1j} \sim (j-1)^{-1/\eta}$ for $0 < V < 2t$.
This behavior is faithfully reproduced by sc-RPA for $-1.0 \lesssim V \lesssim 3.0$ where the iteration is stable. In Fig.3, we compare the correlation function from sc-RPA and bosonization at $V=1.0$, $1.5$, and $-1.0$. Considering the size difference ($L=192$ for sc-RPA and $L=\infty$ for bosonization), the agreement is remarkable. This supports that the self-consistent calculation of sc-RPA indeed captures some major features of the Luttinger liquid ground state. Since the decay exponent of $C_{1j}$ is related to the exponent of $\langle n_k \rangle$ distribution at $k_F$, we have reason to expect that the $\langle n_k \rangle$ from sc-RPA should also obey the Luttinger liquid form in the thermodynamic limit. This issue awaits further investigation.

\begin{figure*}[htbp]
\vspace*{-3.0cm}
 \centering
  \includegraphics[width=550pt, height=420pt, angle=0]{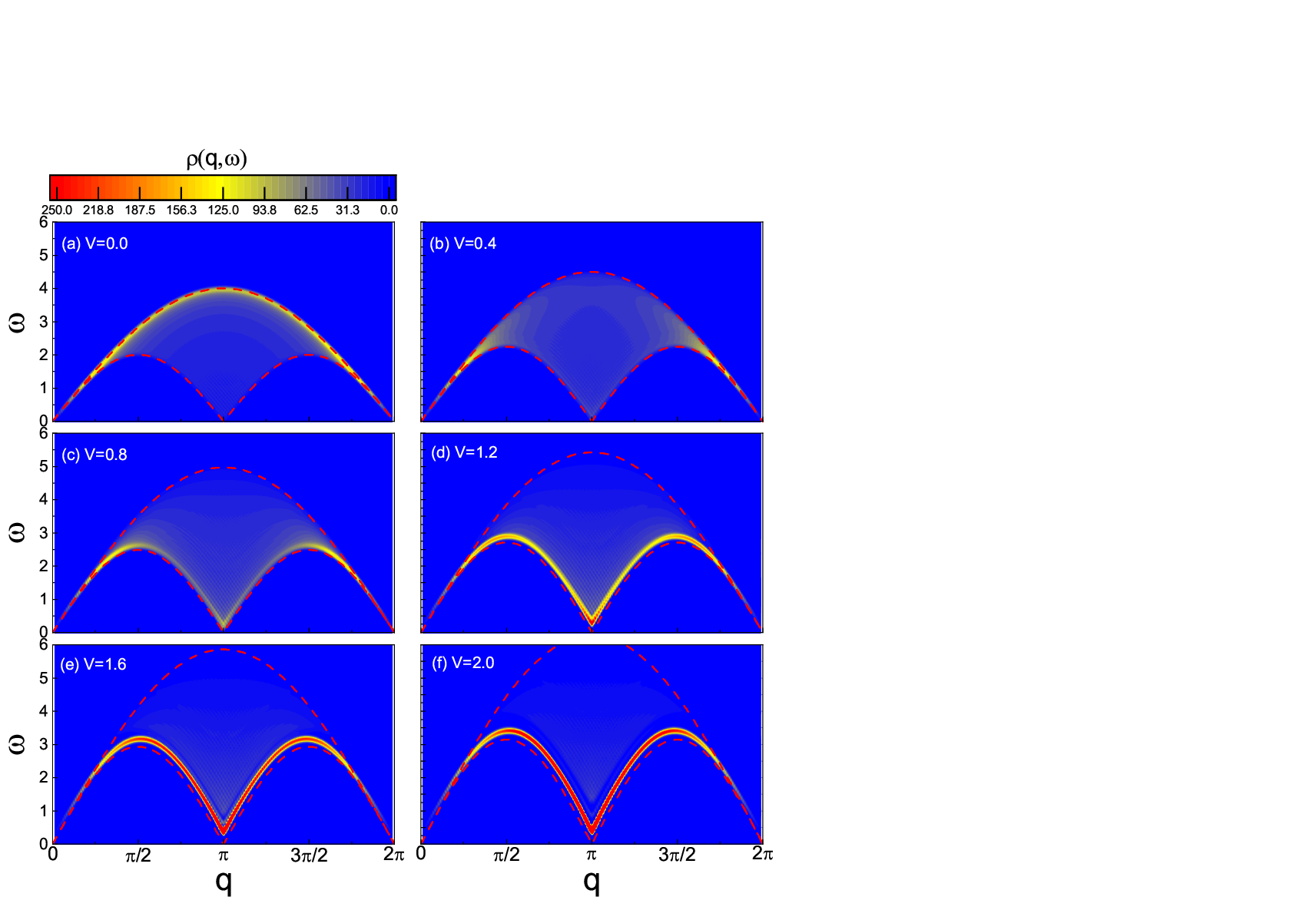}
  \vspace*{-1.0cm}
  \caption{(color online) The color map of the spectral function $\rho(q, \omega)$ for $V>0$. They are obtained at $L=192$, $N=96$, $t=1.0$, $T=0.01$, $\Delta=10^{-5}$, with broadening parameter $\eta=0.05$. The red dashed lines are the boundaries of the particle-hole excitation continuum obtained by Bethe ansatz \cite{Caux1,Caux2}. 
}   \label{Fig4}
\end{figure*}

In Fig.4 and Fig.5, we investigate the density spectral function $\rho(q, \omega) \equiv -(1/\pi) \text{Im} G(\rho_q| \rho_q^{\dagger})_{\omega + i\eta}$, with $\rho_q = \sum_{k} c_{k}^{\dag}c_{k+q}$ and $\eta = 0^{+}$. At zero temperature, this quantity is related to the longitudinal dynamic spin correlation function of XXZ model which has received extensive study in the past decades \cite{Lukyanov1,Hikihara1}. For the repulsive regime $0 < V < 2t$, the excitations are composed of spinon continuum in the frequency window $[\omega_l(q), \omega_u(q)]$, with the exact boundary from Bethe ansatz \cite{Caux1,Caux2}
\begin{eqnarray}\label{Eq7-3}
&& \omega_{l}(q) = v_F |\sin{q}|,  \nonumber \\
&& \omega_{u}(q) = 2v_F \sin{(q/2)},
\end{eqnarray}
where
\begin{eqnarray}\label{Eq7-4}
 v_F=\pi t\frac{\sin(\mu)}{\mu}, \,\,\,\,\,\, \mu=\arccos[V/(2t)].
\end{eqnarray}
Inside the frequency window, the distribution of spectral weight evolves with increasing $V$. At $V=0$, there is a $[\omega_u(q)-\omega]^{-1/2}$ singularity in the spectral function at the upper boundary. This singularity arises from the van-Hove singularity of the single-particle dispersion. sc-RPA is exact at $V=0$, but the exact description of the singular behavior requires the large size limit. As $V$ increases, the spectral weight first moves towards the two wings of the continuum at $q \sim 0$ and $2\pi$. As $V$ increases above $0.8$, the weights further shifts towards the lower boundary. Finally, close to the critical point $V=2.0$, the weights is concentrated to the lower boundary around $q=\pi$, reflecting strong anti-ferromagnetic fluctuation in the spin language. Fig.4 shows that both the boundary position and the qualitative trend in the weight redistribution are well described by the sc-RPA, as compared with the results from two-spinon calculation \cite{Caux1,Caux2}. 

Note that our sc-RPA is carried out with the constrains of the full symmetry of Hamiltonian and hence does not allow symmetry spontaneous breaking. We expect that with relaxed symmetry constraints, the symmetry broken state in $V>2t$ regime can also be captured by sc-RPA calculation.

\begin{figure*}
 \vspace*{-3.0cm}
\centering
  \includegraphics[width=550pt, height=420pt, angle=0]{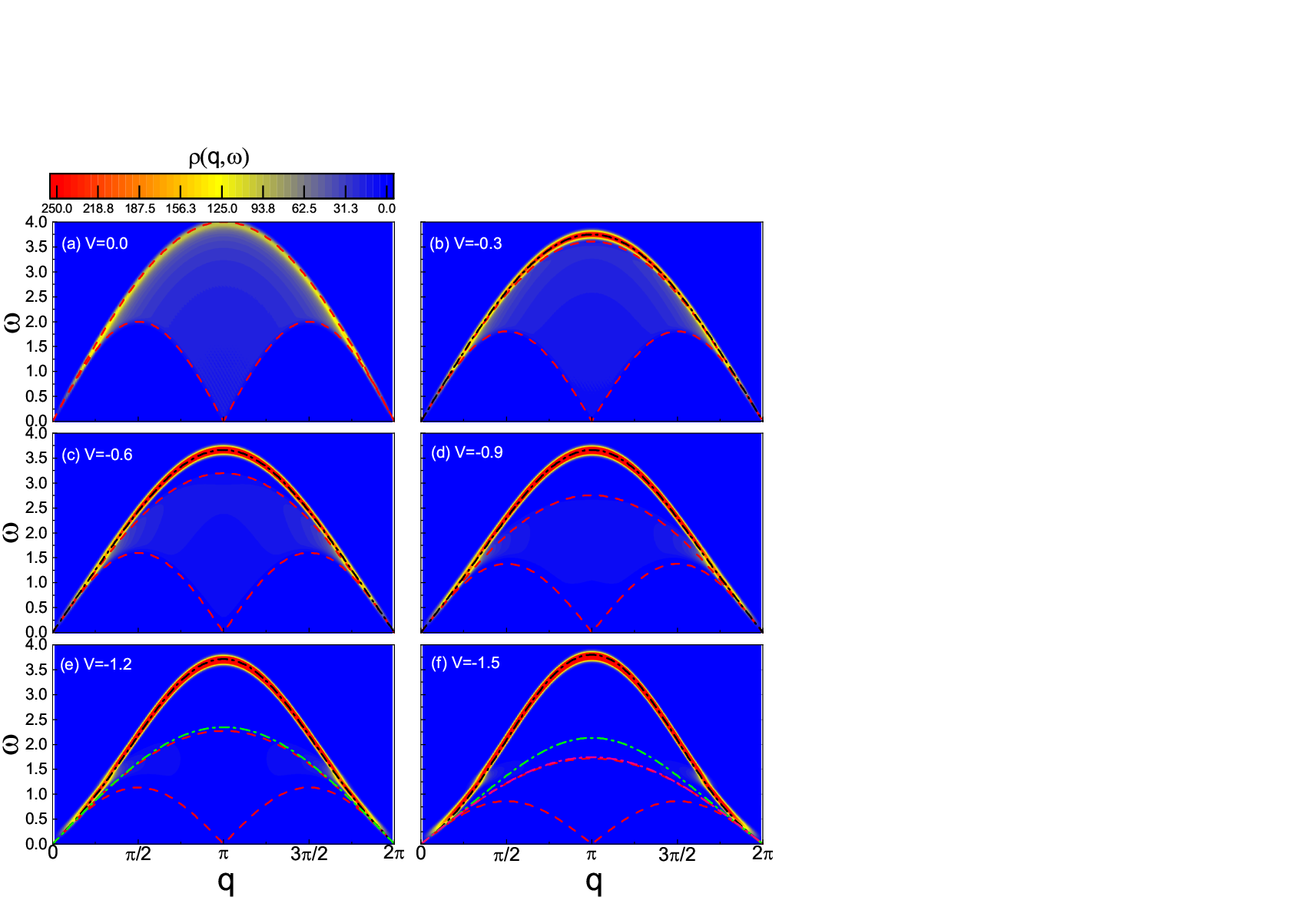}
  \vspace*{-1.0cm}
  \caption{(color online) The color map of the spectral function $\rho(q, \omega)$ for $V<0$. Parameters are same as those of Fig.4. The red dashed lines are the boundaries of the particle-hole excitation continuum obtained by Bethe ansatz \cite{Caux1,Caux2}. The black, green, and pink dot-dashed lines are the bound state dispersions from Eq.(\ref{Eq7-5})-(\ref{Eq7-6}).
}   \label{Fig5}
\end{figure*}

In Fig.5, we present the color map of the spectral function for the regime $-2t < V < 0$. Unlike the regime $0 < V < 2t$, the attractive regime corresponds to the ferromagnetic XXZ model. It is known that the spectral function is featured by a continuum in the frequency window $[\omega_l(q), \omega_u(q)]$ together with possible discrete bound state excitations $\omega^{B}_{n}(q) > \omega_u(q)$ \cite{Schneider1},
\begin{eqnarray}\label{Eq7-5}
 \omega^{B}_n(q) &=& \omega_{u}(q)\frac{\sqrt{1-\cos^{2}(y_{n})\cos^{2}(q/2)}}{\sin(y_{n})},
\end{eqnarray}
where
\begin{eqnarray}\label{Eq7-6}
y_{n}=\frac{n\pi}{2}(\frac{\pi}{\mu}-1)<\frac{\pi}{2}.
\end{eqnarray}
Here, $\mu=\arccos[V/(2t)]$. 
The bound states appear for $V <0$ ($n=1)$, $V<-1$ ($n=2$), and $V<-\sqrt{2}$ ($n=3$), etc. They are plotted in Fig.5 by the black, green, and pink dot-dashed lines in respective panels. As shown in Fig.5, the sc-RPA results still reproduce the accurate continuum boundaries and the first bound state at $\omega^{B}_{1}(q)$. Our results show that as $V$ decreases from zero, the spectral weights in the continuum are quickly concentrated to the first bound state, and only small weights are left in the continuum. At $V=-1.2$ and $V=-1.5$ (panels (e) and (f) of Fig.5) where the second and third bound states appear above $\omega_u(q)$, there are almost no weights left in the continuum and sc-RPA failed to produce the excitations at $\omega^{B}_{2}(q)$ and $\omega^{B}_{3}(q)$, whose exact spectral weights are expected to be very small. 

We also plot the excitation energies as a function of $q$ (not shown here). We find that although the spectral function looks rather satisfactory, the excitation energies in the continuum window have a shift upward with increasing $|V|$, opening a small gap, which cannot be seen in the color map because the weight is very small. The first bound state energy that carries the most spectral weight does not have such a shift and remains very close to the exact value $\omega_1^{B}(q)$.  

In summary, the above results of sc-RPA for the one-dimension spinless fermion model have shown that the ground state and excitations are quantitatively described by sc-RPA in the range $-1.0 \lesssim V \lesssim 3.0$, including the Luttinger-liquid correlation function, particle-hole continuum, and the bound state excitation in $V<0$ regime. Considering the boson nature of the density fluctuations in one dimensional system, one may not be surprised that the particle-hole basis works well. But the self-consistent building of the ground state (described by the one- and two-body reduced density matrices) is vital for the success of sc-RPA. 

For larger $|V|$ values, the sc-RPA calculation becomes unstable and is accompanied with the violation of the negativity of ${\bf L}$. These strong coupling regimes could be better described by sc-RPA with different basis operators or inner product. In this regards, the present PTA-based formalism is a flexible framework that allows RPA to be extended to various situations. Recently, renewed interest emerges in the study of 1D spinless fermion models with finite doping \cite{Schlottmann1,Cha1} or 3-site interactions \cite{Baktay1}, and in 2D systems with \cite{Zhang1} or without \cite{Menczer1} flux. We expect that the present sc-RPA can be used in the study of these frontier problems.

\end{subsection}

\end{section}

\begin{section}{Summary and Discussions}

   In this section, we discuss several issues of the sc-RPA developed based on the PTA formalism.
   
\begin{subsection}{Approximations in RPA}

Here, we summarize the approximations introduced in the various stages of derivation of RPAs.  
First, we believe that the basis $\mathcal{B}_1$ is not a truly complete quadratic basis. The truly complete quadratic basis should also include the number operator $a_{\alpha}^{\dagger}a_{\alpha}$ ($\alpha \in [1, L]$). Although under present inner product definition Eq.(\ref{Eq9}), the length of the number operator  
$\sqrt{|(n_{\alpha}|n_{\alpha})|}$ is zero, there seems to be no reason to believe that $(\hat{n}_{\alpha})_d = 0$ since $[\hat{n}_{\alpha}, H] \neq 0$. That is, $\hat{n}_{\alpha}$ should have dynamics and contribute to the eigen-excitation operators. The truly complete basis $\mathcal{B}_{0} = \{ a_{\alpha}^{\dagger}a_{\beta} \}$ ($1 \leq \alpha, \beta \leq L$) is invariant under orbital rotation. We therefore expect that the PTA on the basis $\mathcal{B}_0$ will give out results that are independent of the orbital selection. However, one may expect that only in the special case of very strong correlation, such as the case with degenerate orbitals, will the dynamic component of $\hat{n}_{\alpha}$ be important. 

From the PTA point of view, we can identify the following approximations made in the RPA formalisms.
\begin{enumerate}
    \item For the sc-RPA on basis $\mathcal{B}_1$, operators higher than the quadratic order are omitted; 
    \item Operators $a_{\alpha}^{\dagger}a_{\alpha}$ ($1 \leq \alpha \leq L$) are omitted from the basis;
    \item The static contribution appearing in the spectral theorem is neglected, i.e., $\langle (a_{\alpha}^{\dagger}a_{\beta})^{\dagger}_0 (a_{\alpha^{\prime}}^{\dagger}a_{\beta^{\prime}})_0 \rangle \approx 0$ ($\alpha \neq \beta$ and $\alpha^{\prime} \neq \beta^{\prime}$). This is exact for $T=0$ and no ground state degeneracy. Otherwise it is not exact;
     \item For PTA on the restricted $\mathcal{B}_2$ basis, higher order terms of the amplitude $Y$ are neglected in the self-consistent calculation of one- and two-particle densities;
     \item In most practical RPA calculations, the working orbitals are pre-selected orbitals and are approximately treated as natural orbitals;
     \item In the renormalized RPA, Hartree-Fock-like decoupling of two-particle densities are introduced ;
     \item In the standard RPA, the Fock contribution in ${\bf L}$ is neglected. The single- and two-particle densities are evaluated on the Hartree-Fock ground state.
\end{enumerate}

\end{subsection}


\begin{subsection}{ The Problem of Singular $\bf{I}$ Matrix }

A notorious problem of the traditional RPA is that it applies only to the symmetry broken phases. Such phases include the ground state where one part of the single particle orbitals are almost fully occupied and the other empty. They also include the magnetic long-range ordered phase where the spins are almost polarized. For the former, $ \langle [ (a_{\alpha}^{\dagger}a_{\beta})^{\dagger}, a_{\alpha}^{\dagger}a_{\beta} ] \rangle = \langle n_{\beta} \rangle -\langle n_{\alpha} \rangle  = \pm 1$ for orbitals $\alpha$ and $\beta$ on the two sides of Fermi surface. For the polarized spin systems, $\langle [S_i^{+}, S_{i}^{-}] \rangle = 2 \langle S_z \rangle = \pm 1$ (for spin 1/2). In both examples, the basis operators of RPA behave more like the well-defined boson operators. They have finite length under the commutator inner product, i.e., $|A_i| = \sqrt{|(A_i|A_i)|} > 0$. As a result, both Rowe's formulation and our sc-RPA are well defined for symmetry broken phases. 
   
In contrast, in the high symmetry states where there are orbitals with very close occupation numbers, $n_{\alpha} \sim n_{\beta}$ (such as the high temperature limit, flat-band fermion systems, or paramagnetic phase of a spin system in the fermion representation), $a_{\alpha}^{\dagger}a_{\beta}$ has very small or zero length and the inner product matrix ${\bf I}_{ij} \sim  \delta_{ij} (\langle n_{\alpha} \rangle - \langle n_{\beta} \rangle) \sim 0$. Both Rowe's formulation and our sc-RPA meet difficulty. In Rowe's formulation, the normalization of basis operators in Eq.(\ref{Eq76}) does not work any more. In PTA-based sc-RPA, the equation ${\bf L} = {\bf I} {\bf M}$ can no longer determine ${\bf M}$ uniquely for a given pair of ${\bf L}$ and ${\bf I}$ due to the singularity of ${\bf I}$. This will manifest itself as the convergence difficulty in the iterative solution of sc-RPA equations. In the practical calculation, our algorithm does not converge for $\Delta=0$ or at high temperatures where there are orbitals with very close occupation number. The convergence of calculation is recovered for $\Delta \geq 10^{-5}$ at low temperature. 
   
   To some extent, the numerical instability can be alleviated by the zero-removing technique \cite{Jia1}. In this technique, one first diagonalizes the inner product matrix $\bf{I}$ and finds all its eigenvalues. The eigenvectors, when combined with the basis operators, produce a set of orthonormal basis operators. From all the $D$ transformed basis operators, we remove the $D_0$ operators that correspond to the $D_0$ smallest absolute eigenvalues. In the remaining $D-D_0$ dimensional space, we carry out a new PTA calculation. In this way, the singular $\bf{I}$ problem is partly solved. In the present work, we did not take this strategy since our sc-RPA works already for $\Delta=10^{-5}$ which is very close to the degeneracy point.

Another possible solution to the problem of singular $\bf{I}$ is to use an inner product other than the commutator one. The commutator inner product $(X|Y) = \langle [X^{\dagger}, Y] \rangle$ used in the existing sc-RPA has the advantage that it has a lower order than both $X$ and $Y$ if they are boson-type operators. But it does not fulfil the mathematical requirement for an inner product that $(X|X) \geq 0$ and $(X|X) = 0$ if and only if $X=0$. Making use of the flexibility of choosing inner product in PTA, one could use other positive inner product to develop the sc-RPA. One could use the anti-commutator inner product to establish RPA, or use the double-commutator inner product $(X|Y) = \langle [X^{\dagger}, [Y, H]] \rangle$ similar to that used for the classical model \cite{Ma1}. This will lead to a sc-RPA theory similar to the spin-rotation invariant GF theory for the short-range order of the Heisenberg model \cite{Winterfeldt1,Ihle1}.

Both the above two inner products are positive definite. In the sc-RPA with these inner product, the zero-length problem of the operators $a_{\alpha}^{\dagger}a_{\alpha}$ no longer exists. It is also possible to add $\{ a_{\alpha}^{\dagger}a_{\alpha} \}$ into the basis and still have a non-singular $\bf{I}$. The basis is then invariant under the orbital rotation. The cost of such schemes is the more complicated $\bf{I}$ and $\bf{L}$ matrices. Note that the static component problem still exists in PTA with these inner products. This direction will be explored in the future. 
  
\end{subsection}

\begin{subsection}{ The Problem of Static Component of Basis Operators    }

  Here, we discuss the problem of the static component of basis operators $\{ A_i \}$ in PTA. The GF EOM method is based on the truncation
\begin{equation}
   [A_i, H]  \approx \sum_{j} M_{ji} (A_{j})_d.    \label{Eq113}
\end{equation}
Since $[A_i, H]_0 = 0$, only the dynamic component of $A_j$ appears on the right-hand side of the above equation. Consequently, PTA can only produce self-consistent equations for the averages of the type $\langle (A_i^{\dagger})_d (A_j)_d \rangle$. Considering that $(A_i)_0$ is in general nonzero and unknown, usually one uses the approximation Eq.(\ref{Eq47}) to get $\langle A_i^{\dagger} A_j \rangle$. For examples, for the general Hamiltonian Eq.(\ref{Eq58}), $(a_{\alpha}^{\dagger} a_{\beta})_0$ is nonzero in general. For the one-dimensional spinless fermion model Hamiltonian Eq.(\ref{Eq6-1}), $(c_{k}^{\dagger} c_{k+q})_0 \neq 0$ (for $q \neq 0$) if two degenerate eigenstates differ in the total momentum by $q$. $c_{k}^{\dagger} c_k$ also has nonzero static component in general. In our sc-RPA derivation, we use the approximation $(c_{k}^{\dagger} c_{k+q})_0 \approx 0$ (for $q \neq 0$) and omit $c_{k}^{\dagger} c_{k}$ in the  operator basis. The numerical results confirm that these approximations have little influence on the accuracy of results in the small to intermediate $V$ regime, but in general the influence may be nonnegligible. Since the static component problem is independent of the definition of inner product, it is a more fundamental problem of PTA.  
Below we discuss several possible ways of selecting basis to overcome this problem in the practical PTA calculation.  

In the ideal case, we can choose the basis operator $A_i$ such that $(A_{i})_0$ is known to be zero. This can be realized in certain situations where the symmetry of $A_i$ and $H$ can guarantee this. For an example, for fermion systems that conserves the total fermion number, we have $(c_{i})_0=0$ for the annihilation operator. 

For systems with no such symmetry, the operators of the form $A_i = [\tilde{A}_i, X]$ have the exact properties $(A_{i})_0=0$. Here $\tilde{A}_i$ is an arbitrary operator and $X$ is a conserved quantity fulfilling $[X, H]=0$. One can choose such operators $A_i$ as the basis to avoid the static component problem. 

Finally, one can choose the basis $A_i$ that has approximately zero static component, {\it i.e.}, $(A_{i})_0 \approx 0$. For an example, one can choose $A_{i} = \tilde{A}_i - \langle \tilde{A}_i \rangle$, which fulfils only the necessary condition $\langle A_{i} \rangle =0$ for the zero static component properties $ (A_{i})_0 =0$. One could also use the eigen-excitation operator $O_{\nu}$ of a low-order PTA calculation to construct $A_i$, such as $A_{i} = O_{\mu}^{\dagger} O_{\nu}$ with $\lambda_{\mu} \neq \lambda_{\nu}$. It also has $(A_{i})_0 \approx 0$.  
\end{subsection}

In summary, in this paper, we derive the sc-RPA in the PTA formalism which is a more systematic and flexible framework.
The present sc-RPA applies to arbitrary temperature. It recovers Rowe's sc-RPA formalism at zero temperature.
We apply the present sc-RPA to one-dimensional spinless fermion model. The N-representability constraints for 2RMD are enforced to significantly stabilize the iterative solution. The Luttinger liquid ground state and the continuum/bound state features of the density spectral function are well captured by the present sc-RPA. 
From the derivation, we rationalize Rowe's formalism and elucidate the approximations contained in various RPAs. The PTA framework also provides new degrees of freedom to extend RPA, i.e., definitions of inner product, selection of basis operators, and constraints of reduced density matrices. These will help people extend RPA to situations of high temperature, higher order correlation, and high symmetry states. Although our formalism is derived for lattice model Hamiltonians, it also paves the way to sc-RPA's application in first principle calculations of real materials.

\end{section}

\begin{section}{Acknowledgments }
N.H.T. acknowledges helpful discussions with C.F. Wang, L. Wan, Y. Wan, Y. Zhong, and T. Li. This work is supported by National Natural Science Foundation of China (Grant Nos. 11974420, 12374067, and 12188101), as well as by the National Key Research and Development Program of China (Grant Nos. 2022YFA1403800 and 2023YFA1507004).
\end{section}


\appendix{}

\section{Proof of Spectral Theorem Eqs.(\ref{Eq4}) and (\ref{Eq5}) }

In this Appendix, we prove the spectral theorem Eqs.(4) and (5), which applies to the Boson-type GF.
The proof involves some properties of the static as well as dynamic components of an operator.

We start from the definition of Boson-type retarded GF for operators $A$ and $B$, Eqs.(1) and (2),
\begin{equation}
   G^{r}[A(t)|B(t^{\prime})] \equiv \frac{1}{i} \theta(t-t^{\prime}) \langle \left[A(t), B(t^{\prime}) \right]\rangle,   \label{EqA1}
\end{equation}
and
\begin{equation}
   G^{r}(A|B)_{\omega} = \int_{-\infty}^{\infty}  G^{r}[A(t)|B(t^{\prime})] e^{i (\omega + i\eta)(t-t^{\prime}) } d(t-t^{\prime}).     \label{EqA2}
\end{equation}
The Lehmann representation for $G^{r}[A(t)|B(t^{\prime})]$ reads
\begin{eqnarray}
  G^{r}[A(t)|B(t^{\prime})] &=& \frac{1}{iZ} \theta (t-t^{\prime}) \sum_{mn} \left( e^{-\beta E_{m}} - e^{-\beta E_n} \right) \times  \nonumber \\
  && \langle m|A|n\rangle\langle n |B|m\rangle e^{i (E_m -E_n) (t-t^{\prime}) }.  \nonumber \\
  &&     \label{EqA3}
\end{eqnarray}
Here, $Z$ is the partition function. $|m \rangle$ and $E_{m}$ are the eigen state and eigen energy of $H$, respectively. Note that here we have used ${\bf 1} = \sum_{m} |m \rangle \langle m|$ in the full Fock space. So Eq.(\ref{EqA3}) is valid in general only for the grand canonical ensemble. For the canonical ensemble, Eq.(\ref{EqA3}) is valid only for the case $[H, N]=[A, N]=[B, N]=0$.  According to Eq.(\ref{Eq6}), we can decompose $A$ and $B$ as $A = A_{0} + A_{d}$, $B = B_{0} + B_d$. Inserting them into the Lehmann representation, one easily obtains
\begin{equation}
   G^{r}[A(t)|B(t^{\prime})] = G^{r}[A_d(t)|B_d (t^{\prime})].     \label{EqA4}
\end{equation}
Similarly, it is easy to prove that $G^{r}[A_{0}(t)|B(t^{\prime})] = G^{r}[A(t)|B_{0}(t^{\prime})]=G^{r}[A_{0}(t)|B_{0}(t^{\prime})]=0$. After Fourier transformation, one obtains
\begin{eqnarray}
  && G(A|B_0)_{\omega} = G(A_0|B)_{\omega} = G(A_0|B_0)_{\omega} =0,  \nonumber \\
  && G(A|B)_{\omega} = G(A_d | B_d)_{\omega}.     \label{EqA5}
\end{eqnarray}

From the Lehmann representation, the average value of commutator $\langle [A, B]\rangle$ has similar relations,
\begin{eqnarray}
&& \langle [A, B_0]\rangle = \langle [A_0, B]\rangle = \langle [A_0, B_0]\rangle =0, \nonumber \\
&& \langle [A, B]\rangle = \langle [A_d, B_d]\rangle.     \label{EqA6}
\end{eqnarray}
Using the above properties, the EOM of GF Eq.(\ref{Eq3}) becomes
\begin{eqnarray}
  \omega   G(A_d|B_d)_{\omega} &=& \langle \left[A_d, B_d\right] \rangle +  G([A_d,H]|B_d)_{\omega}   \nonumber \\
  &=& \langle \left[A_d, B_d\right] \rangle -  G(A| [B_d, H])_{\omega} .     \label{EqA7}
\end{eqnarray}
This shows that only the dynamic component of operators are involved in the Boson-type GF and its EOM.

From Eq.(A5), one has the spectral function $\rho_{A,B}(\omega) = \rho_{A_d, B_d}(\omega)$.
Lehmann representation of  $\rho_{A,B}(\omega)$ then reads
\begin{eqnarray}
  \rho_{A,B}(\omega) &=& \frac{1}{Z} \sum_{(mn)^{\prime}}\left( e^{-\beta E_{m}} - e^{-\beta E_n} \right) \times \nonumber \\
   &&  \langle m|A|n\rangle\langle n |B|m\rangle \delta(\omega + E_m - E_n).      \label{EqA8}
\end{eqnarray}
Here, $\sum_{(mn)^{\prime}}$ means summation for $m$ and $n$ with the constraint $E_m \neq E_n$.
From Eq.(\ref{EqA8}), one obtains
\begin{eqnarray}
  \int_{-\infty}^{\infty} \rho_{A,B}(\omega) \frac{1}{e^{\beta \omega} -1} &=& \frac{1}{Z} \sum_{(mn)^{\prime}} \langle m |A|n\rangle \langle n|B| m \rangle e^{-\beta E_{n}}  \nonumber \\
&=& \langle B_d A_d \rangle = \langle B A \rangle -\langle B_0 A_0 \rangle . \nonumber \\
&&     \label{EqA10}
\end{eqnarray}
This completes the proof of Eqs.(4) and (5).

\section{The Applicability of Spectral Theorem in Canonical / Grand Canonical Ensemble}

For a given Hamiltonian $H$ that conserves the total number of electrons, i.e., $[H, N]=0$,
by writing down the EOM for $G(A|B)_{\omega}$, truncating the hierarchy of EOMs, and using spectral theorem, one can establish approximate relations between the average $\langle BA \rangle$ and others such as $\langle [A,B] \rangle$ (for the Boson-type GF). In this appendix, we will prove that if $A$ and $B$ conserves the total number of electrons,
i.e., $[H,N] = [A, N] = [B, N] = 0$, those relations will be applicable to both the canonical ensemble and the grand canonical ensemble. Otherwise they apply only to the grand canonical ensemble.

One simple example is a single mode model, $H=(\epsilon -\mu) a^{\dagger}a$. For the one-particle GF $G(a|a^{\dagger})_{\omega}$, which has $[a, N] \neq 0$, the associated spectral theorem gives $\langle a^{\dagger}a \rangle = 1/ [e^{\beta (\epsilon-\mu)} - 1 ]$. This expression obviously applies to the grand canonical ensemble, but not to the canonical ensemble. We note that the distinction between canonical and grand canonical ensemble only exist at $T > 0$. At $T=0$, there is no distinction if the ground state has a definite number of electrons.

Here, we only consider the Boson-type GF. Losing no generality, we assume that $A$ is an eigen excitation operator $[A, H] = \epsilon A$ and $A_0 = 0$. 
Writing down the GF EOM for $G(A|B)_{\omega}$ and applying the standard spectral theorem, Eq.(\ref{Eq4}), we obtain
\begin{equation}
  \langle BA \rangle = \frac{ \langle [A, B] \rangle}{e^{\beta \epsilon} -1}.     \label{EqB1}
\end{equation}

Below, we first prove that if $[H,N] = [A, N]=[B,N]=0$, Eq.(\ref{EqB1}) applies both to the canonical and the grand canonical ensemble.  
In the canonical ensemble, the Lehmann representation for the left-hand side of this equation reads
\begin{eqnarray}
 && \langle BA \rangle_{N} \nonumber \\
 &=& \frac{1}{ {\text Tr}_{N} \left[ e^{-\beta H} \right]} {\text Tr}_{N} \left[e^{-\beta H} BA \right] \nonumber \\
  &=& \frac{1}{\sum_{n} e^{-\beta E_{n}(N)} } \sum_{(mn)^{\prime}}e^{-\beta E_{n}(N)} \langle N, n |B|N, m\rangle \langle N,m | A | N,n\rangle. \nonumber \\
  &&     \label{EqB2}
\end{eqnarray} 
Here, $\sum_{(mn)^{\prime}}$ means summation for $m$ and $n$ with the constraint $E_m \neq E_n$. $|N,n\rangle$ and $E_{n}(N)$ are eigen state and energy of $H$ in the subspace $N$, respectively. $\langle ... \rangle_N$ is the canonical ensemble average in the $N$-electron subspace. $[H,N]=[A, H]=[B,H]=0$ has been used in the above equation.

For the right-hand side of Eq.(\ref{EqB1}), we have
\begin{eqnarray}
 &&  \langle [A, B]\rangle_{N} \nonumber \\
 &=& \frac{1}{\sum_{n} e^{-\beta E_{n}(N)} } \sum_{(mn)^{\prime}} e^{-\beta E_{n}(N)}  \times \nonumber \\
 && \left[e^{\beta [E_{n}(N) - E_{m}(N)]} -1 \right] \langle N, n |B|N, m\rangle \langle N,m | A | N,n\rangle. \nonumber \\
 &&     \label{EqB3}
\end{eqnarray}
Sandwiching $[A, H] = \epsilon A$ in the eigen states $\langle N,m |$ and $|N, n\rangle$, one obtains $\left[E_{n}(N) - E_{m}(N) \right] \langle N,m | A |N,n \rangle = \epsilon\langle N,m | A |N, n\rangle$, which further gives $e^{\beta \left[E_{n}(N) - E_{m}(N) \right]} \langle N,m | A |N, n \rangle = e^{\beta \epsilon}\langle N,m | A |N, n\rangle$. Putting this relation into Eq.(\ref{EqB3}) gives
\begin{equation}
  \langle BA \rangle_N = \frac{ \langle [A, B] \rangle_N }{e^{\beta \epsilon} -1}.     \label{EqB4}
\end{equation}
That is, Eq.(\ref{EqB1}) holds for the canonical ensemble if $[H,N]= [A, N]=[B,N] =0$.

In the grand canonical ensemble, we replace $H$ with $H-\mu N$. Now $[A, H-\mu N] = [A, H] = \epsilon A$. The left-hand side of Eq.(\ref{EqB1}) reads
\begin{eqnarray}
 \langle BA \rangle_{G} &=& \frac{ {\text Tr} \left[e^{-\beta (H-\mu N)} BA \right]}{ {\text Tr} \left[ e^{-\beta (H-\mu N)} \right]}  \nonumber \\
 &=& \frac{ \sum_{N} e^{\beta \mu N } Z_N \langle BA \rangle_N  }{\sum_{N} e^{\beta \mu N} Z_N}.      \label{EqB5}
\end{eqnarray}
Here $Z_N = \sum_{n} e^{-\beta E_{n}(N)}$ is the partition function in the $N$-particle space. 

The right-hand side of Eq.(\ref{EqB1}) reads
\begin{eqnarray}
 && \frac{1}{e^{\beta \epsilon} -1}  \langle [A, B] \rangle_{G} \nonumber \\
 && = \frac{1}{e^{\beta \epsilon} -1}  \frac{ \sum_{N} e^{\beta \mu N } Z_N \langle [A,B] \rangle_N  }{\sum_{N} e^{\beta \mu N} Z_N}.      \label{EqB6}
\end{eqnarray}
Employing Eq.(\ref{EqB4}) and comparing the results with the left-hand side Eq.(\ref{EqB5}), one arrives at 
\begin{equation}
  \langle BA \rangle_G = \frac{ \langle [A, B] \rangle_G }{e^{\beta \epsilon} -1}.     \label{EqB7}
\end{equation}
Therefore, we have proved that Eq.(\ref{EqB1}) applies to both canonical and grand canonical ensemble if $[H,N]= [A, N]=[B,N] =0$.

Now let us prove that if $[A, N] \neq 0$, Eq.(\ref{EqB1}) does not hold for the canonical ensemble, but still holds for the grand canonical ensemble.
Suppose $[A, N] = q A$ and $q \neq 0$. For the canonical ensemble, we have
\begin{eqnarray}
 && \langle BA \rangle_{N} \nonumber \\
 &=& \frac{1}{Z_N } \sum_{nm}e^{-\beta E_{n}(N)} \langle N, n |B|N-q, m\rangle \langle N-q,m | A | N,n\rangle, \nonumber \\
  &&     \label{EqB8}
\end{eqnarray} 
and
\begin{eqnarray}
 &&  \langle [A, B]\rangle_{N} \nonumber \\
 &=& \frac{1}{Z_N} \sum_{nm} e^{-\beta E_{n}(N)}  [ \langle N, n |A|N+q, m\rangle \langle N+q,m | B | N,n\rangle   \nonumber \\
 && - \langle N, n |B|N-q, m\rangle \langle N-q,m | A | N,n\rangle   ]. \label{EqB9}
\end{eqnarray}
Using $[A, H] = \epsilon A$, we obtain
\begin{eqnarray}
 &&  \langle [A, B]\rangle_{N} \nonumber \\
 &=& \frac{e^{\beta \epsilon}}{Z_N}  \sum_{nm} e^{-\beta E_{m}(N+q)}  \langle N, n |A|N+q, m\rangle \langle N+q,m | B | N,n\rangle   \nonumber \\
 && - \frac{1}{Z_N} \sum_{nm} e^{-\beta E_{n}(N)}  \langle N, n |B|N-q, m\rangle \langle N-q,m | A | N,n\rangle    \nonumber \\
&& \neq (e^{\beta \epsilon} -1) \langle BA \rangle_{N}.  \label{EqB10}
\end{eqnarray}
Therefore, Eq.(\ref{EqB4}) does not hold.

For the grand canonical ensemble, we replace $H$ with $H-\mu N$ and $\epsilon$ is now defined by $[A, H-\mu N] = \epsilon A$. So $E_{m}(N+q) - E_{n}(N) = \epsilon + \mu q$ if $\langle N,n | A |N+q,m\rangle \neq 0$. We put this equation into the expression Eq.(\ref{EqB9}) for $\langle [A,B] \rangle_N$,  do the replacement $N \rightarrow N-q$ and exchange $m$ and $n$ in the summation of the first term. Inserting the obtained equation into Eq.(\ref{EqB6}) for$\langle [A,B] \rangle_G$ and comparing it with Eq.(\ref{EqB5}) for $\langle BA \rangle_G$, we obtain the grand canonical equation Eq.(\ref{EqB7}). This completes the proof that Eq.(\ref{EqB7}) holds in general while Eq.(\ref{EqB4}) holds only for the case $[H,N] = [A,N]=[B,N]=0$. For the operators $A$ and $B$ with nonzero static components, $A_0 \neq 0$ and $B_0 \neq 0$, the above proof is still valid if we replace $A$ with $A_d$ and $B$ with $B_d$.

\section{Negative Semi-definiteness of Liouville Matrix {\bf L}}

In this Appendix, we will prove that the Liouville matrix ${\bf L}$ defined in Eq.(\ref{Eq10}) is negative semi-definite.
$\bf{L}$ is defined as $L_{ij} = (A_i|[A_j, H])$ with the inner product $(X|Y) \equiv \langle [X^{\dagger}, Y] \rangle$.
From Eq.(\ref{Eq20p}), we can diagonalize ${\bf L}$ as ${\bf U}^{\dagger} {\bf L} {\bf U} = \tilde{{\bf L}}$. The diagonal element of $\tilde{\bf{L} }$ reads 
\begin{equation}
  \tilde{L}_{ii} = (O_i|[O_i, H]) = \langle [O_{i}^{\dagger}, [O_i, H] ] \rangle.       \label{EqC1}
\end{equation}
Here $O_i = \sum_{j} U_{ji} A_j$ is the transformed operator.
The Lehmann representation of $\tilde{L}_{ii}$ reads
\begin{eqnarray}
  \tilde{L}_{ii} &=& \frac{1}{Z} \sum_{mn} e^{-\beta E_m}  \times  \nonumber \\
 && \left[ \langle m|O_{i}^{\dagger}|n\rangle \langle n|[O_i, H]|m \rangle 
 - \langle m| [O_{i}, H]|n\rangle \langle n|O_i^{\dagger}|m\rangle \right] \nonumber \\
 &=& \frac{1}{Z} \sum_{mn}\langle m| O_{i}^{\dagger}|n\rangle \langle n|O_i|m\rangle \left(e^{-\beta E_m} - e^{-\beta E_n} \right) (E_m - E_n) \nonumber \\
 & \leq & 0.
\end{eqnarray}     \label{EqC2}
This completes the proof.

\section{Derivation of Eqs.(\ref{Eq73}) and (\ref{Eq74}) }

In this appendix, we prove Eqs.(\ref{Eq73}) and (\ref{Eq74}) using the number operator method.
In the canonical ensemble, for the system with $N$ electrons, we have $\langle a_{\alpha}^{\dagger} \hat{N} a_{\beta} \rangle = (N-1) \langle a_{\alpha}^{\dagger}a_{\beta} \rangle$.
Using $\hat{N} = \sum_{\gamma=1}^{L} a_{\gamma}^{\dagger} a_{\gamma}$, for $\alpha \neq \beta$ one has 
\begin{equation}
 (N-1) \langle a_{\alpha}^{\dagger}a_{\beta} \rangle = \sum_{\gamma \neq \alpha, \beta} \langle a_{\alpha}^{\dagger}a_{\gamma}^{\dagger} a_{\gamma} a_{\beta} \rangle.     \label{EqD1}
\end{equation}
Replacing $a_{\gamma}^{\dagger}a_{\gamma}$ by $1 -a_{\gamma}a_{\gamma}^{\dagger} $ in the above equation, one obtains
\begin{eqnarray}
 (N-1) \langle a_{\alpha}^{\dagger}a_{\beta} \rangle = (L-2) \langle a_{\alpha}^{\dagger}a_{\beta} \rangle - \sum_{\gamma \neq \alpha, \beta} \langle  a_{\alpha}^{\dagger} a_{\gamma} a_{\gamma}^{\dagger} a_{\beta} \rangle, \nonumber \\
 &&     \label{EqD2}
\end{eqnarray}
which gives Eq.(\ref{Eq73}).

For $\alpha = \beta$, instead of Eq.(\ref{EqD1}), one has
\begin{equation}
 (N-1) \langle a_{\alpha}^{\dagger}a_{\alpha} \rangle = \sum_{\gamma \neq \alpha} \langle a_{\alpha}^{\dagger}a_{\gamma}^{\dagger} a_{\gamma} a_{\alpha} \rangle.        \label{EqD3}
\end{equation}
Using the same method as above, this produces
\begin{eqnarray}
 (N-1) \langle a_{\alpha}^{\dagger}a_{\alpha} \rangle = (L-1) \langle a_{\alpha}^{\dagger}a_{\alpha} \rangle - \sum_{\gamma \neq \alpha} \langle  a_{\alpha}^{\dagger} a_{\gamma} a_{\gamma}^{\dagger} a_{\alpha} \rangle, \nonumber \\
 &&     \label{EqD4}
\end{eqnarray}
which gives Eq.(\ref{Eq74}).

\section{Algorithms for solving generalized eigen problem and linear equations}

In this Appendix, we summarize the two algorithms used in our implementation of sc-RPA for the one-dimensional spinless fermion model. 

\begin{subsection}{generalized eigen problem}

First, we summarize the algorithm of solving the generalized eigen problem Eq.(\ref{Eq6-18.5}), i.e.,
\begin{equation}       \label{EqE1}
 {\bf L}^{q} {\bf U}^{q} = {\bf I}^{q} {\bf U}^{q} {\bf \Lambda}^{q}.
\end{equation}
We seek for the matrices ${\bf U}^{q} $ and the diagonal ${\bf \Lambda}^{q} $ for a given pair of Hermitian matrices ${\bf L}^{q} \preceq 0$ and ${\bf I}^{q} $. 
Since the same algorithm applies to every $q$, below, we drop the superscript $q$.
 Given that ${\bf L}$ is negative semi-definite and Hermitian, we use the following two-step diagonalization method to find ${\bf U}$ and ${\bf \Lambda}$. We first seek for the unitary matrix ${\bf U}_{L}$ (${\bf U}_{L}^{\dag} = {\bf U}_{L}^{-1}$) to diagonalize ${\bf L}$,
\begin{equation}       \label{EqE2}
  ({\bf U}_{L} )^{-1} \, {\bf L}  \, {\bf U}_{L} = {\bf \Lambda}_{L}
\end{equation}
${\bf \Lambda}_{L}$ is a diagonal and negative semi-definite matrix. We can normalize it to $-{\bf 1}$ by
\begin{equation}       \label{EqE3}
|{\bf \Lambda}_{L}|^{-1/2}  ({\bf U}_{L} )^{-1}  \, {\bf L} \,  {\bf U}_{L} |{\bf \Lambda}_{L}|^{-1/2} = - {\bf 1}.
\end{equation}
Carrying out the same transformation to ${\bf I}$, we have
\begin{equation}       \label{EqE4}
|{\bf \Lambda}_{L}|^{-1/2}  ({\bf U}_{L} )^{-1}  \, {\bf I}  \,  {\bf U}_{L} |{\bf \Lambda}_{L}|^{-1/2} = \tilde{{\bf I}},
\end{equation}
which is an Hermitian matrix. Now we seek another unitary transformation ${\bf V}$ (${\bf V}^{\dag} = {\bf V}^{-1}$) to diagoanlize $\tilde{{\bf I}}$,
\begin{equation}       \label{EqE5}
 {\bf V}^{-1}  \, \tilde{{\bf I}} \,  {\bf V} = \tilde{{\bf \Lambda}}_{I}.
\end{equation}
with $\tilde{{\bf \Lambda}}_{I}$ being diagonal matrix.
Now we get the combined transformation matrix ${\bf U} = {\bf U}_{L} |{\bf \Lambda}_{L}|^{-1/2} {\bf V}$ such that
\begin{eqnarray}       \label{EqE6}
&&  {\bf U}^{\dag} {\bf L} {\bf U} = - {\bf 1},  \nonumber \\
&&  {\bf U}^{\dag} {\bf I} {\bf U} = {\tilde{\bf \Lambda}}_{I}.
\end{eqnarray}
Comparing them with Eq.(\ref{EqE1}), we identify that
\begin{eqnarray}       \label{EqE7}
&&  {\bf U} = {\bf U}_{L} |{\bf \Lambda}_{L}|^{-1/2} {\bf V},  \nonumber \\
&&  {\bf \Lambda} = -( {\tilde{\bf \Lambda}}_{I})^{-1}.
\end{eqnarray}
It is easy to prove that the same transformation ${\bf U}$ diagonalized both ${\bf M}$ and ${\bf C}$ as
\begin{eqnarray}       \label{EqE8}
&&  {\bf U}^{-1} {\bf M} {\bf U} = {\bf \Lambda},  \nonumber \\
&&  {\bf U}^{\dag} {\bf C} {\bf U} = {\bf \Lambda}^{-1} \left( e^{\beta {\bf \Lambda}} - {\bf 1} \right)^{-1}.
\end{eqnarray}

Once the generalized eigen vector ${\bf U}^{q}$ and eigen values ${\bf \Lambda}^{q}$ are obtained, one can proceed to 
produce $[e^{\beta {\bf M}^{q}}  - {\bf 1}]^{-1}$, which is used in the fluctuation-dissipation theorem to calculate ${\bf C}^{q}$.

\end{subsection}

\begin{subsection}{linear equations for ${\bf C}^{q}$ and $\{ \langle n_k \rangle \}$ }

Second, we summarize the algorithm of solving the linear equations for ${\bf C}^{q}$ and $\{ \langle n_k \rangle\}$. The set of linear equations to be solved for a fixed ${\bf M}^{q}$ are Eqs.(\ref{Eq6-7}), (\ref{Eq6-9}), and (\ref{Eq6-11}). That is,
\begin{eqnarray}       \label{EqE9}
&& {\bf C }^{q} = - {\bf I}^{q} \left( e^{\beta {\bf M}^{q} } - \bf{1} \right)^{-1}, \nonumber \\
&& \langle n_{k} \rangle = \frac{1}{L-N} \sum_{k^{\prime} (\neq k)} C^{k - k^{\prime}}_{k^{\prime}k^{\prime}}, \nonumber \\
&&  \sum_k \langle n_k \rangle = N.
\end{eqnarray}
We denote ${\bf W}^{q} = \left( e^{\beta {\bf M}^{q} } - \bf{1} \right)^{-1} $. Inserting the expression for $C^{q}_{k, k^{\prime}}$ of the first equation into the second one, and using $I^{q}_{kk^{\prime}} = \delta_{k k^{\prime}} \left( \langle n_{k+q} \rangle - \langle n_k \rangle \right)$, we obtain the linear equations for $\langle n_k \rangle$ as 
\begin{equation}       \label{EqE10}
    \sum_{k^{\prime}} Q_{k k^{\prime}} \langle n_{k^{\prime}} \rangle = \langle n_k \rangle.
\end{equation}
The matrix ${\bf Q}$ is given by
\begin{equation}       \label{EqE11}
  Q_{k k^{\prime}} = \left\{
     \begin{array}{lll} 
        \frac{1}{L-N} W^{k-k^{\prime}}_{k^{\prime} k^{\prime}},  & \,\,\, (k \neq k^{\prime}) \\
        && \\
        \frac{-1}{L-N} \sum_{p} (\neq k) W^{k-p}_{p p}, & \,\,\, (k = k^{\prime}).
     \end{array}\right.    
\end{equation}
We solve the homogeneous linear equations Eq.(\ref{EqE10}) for $\{ \langle n_k \rangle \}$ under the constraints $\sum_k \langle n_k \rangle = N$ by minimizing the distance $d = \left[ ({\bf Q}- {\bf 1} ) \vec{x} \right]^{T} \cdot \left[ ( {\bf Q}- {\bf 1} ) \vec{x} \right] - \lambda (\sum_{k} x_k - N)$ with respect to $\vec{x}$. Here $x_k = \langle n_k \rangle$ are the unknown variables and $\lambda$ is the Lagrangian multiplier to enforce the constraint of the total number of fermions.
This gives an extended $N+1$ dimensional inhomogeneous linear equation that we solve for $\langle n_k \rangle$,
\begin{equation}       \label{EqE12}
\left(
\begin{array}{cc}
2 ({\bf Q} - {\bf 1})^{T} ({\bf Q} - {\bf 1} )  & - \vec{1}  \\
- \vec{1}^{T} & 0
\end{array}  \right)
\left(
\begin{array}{c} 
\vec{x} \\
\lambda
\end{array} \right)
=
\left(
\begin{array}{c} 
\vec{0} \\
-N
\end{array}
\right).
\end{equation}
Here, $\vec{1} = (1 \, 1\, ... \, 1)^{T}$ and $\vec{0} = (0 \, 0 \, ... \, 0)^{T}$.
After $\{ \langle n_k \rangle \}$ are obtained from this equation, we produce ${\bf I}^{q}$ and then calculate the matrix ${\bf C}^{q}$ via the first equation of Eq.(\ref{EqE9}).

\end{subsection}


\end{document}